\documentclass[11pt]{article}

\usepackage[a4paper,margin=1.05in,includefoot]{geometry}
\usepackage[T1]{fontenc}
\usepackage[utf8]{inputenc}
\usepackage{lmodern}
\usepackage{microtype}
\usepackage{amsmath,amssymb,amsthm,mathtools}
\usepackage{booktabs}
\usepackage{array}
\usepackage{longtable}
\usepackage{enumitem}
\usepackage{needspace}
\usepackage{placeins}
\usepackage[hidelinks]{hyperref}
\ifdefined\pdfinfoomitdate
\fi
\newtheorem{theorem}{Theorem}[section]
\newtheorem{lemma}[theorem]{Lemma}
\newtheorem{proposition}[theorem]{Proposition}
\newtheorem{corollary}[theorem]{Corollary}
\theoremstyle{definition}
\newtheorem{definition}[theorem]{Definition}
\newtheorem{example}[theorem]{Example}
\theoremstyle{remark}
\newtheorem{remark}[theorem]{Remark}

\newcommand{\wt}{\operatorname{wt}}
\newcommand{\AT}{d_{\mathrm{AT}}}
\newcommand{\F}{\mathbb{F}}

\newcommand{\cC}{\mathcal{C}}
\newcommand{\cD}{\mathcal{D}}

\hypersetup{
  pdftitle={Concatenation Locality under Successive Adjacent Transpositions},
  pdfauthor={Adelin Ataba; Aryeh Lev Zabokritskiy (Yohananov)},
  pdfsubject={Correction under a bounded number of successive adjacent transpositions},
  pdfkeywords={adjacent transpositions, block concatenation, error correction, streaming decoding, decoding delay, finite automata},
  pdflang={en-US},
  pdfdisplaydoctitle=true
}
\title{Concatenation Locality under\\
Successive Adjacent Transpositions}
\author{Adelin Ataba\\
\normalsize Department of Computer Science\\
\normalsize Technion---Israel Institute of Technology\\
\normalsize ORCID: \href{https://orcid.org/0009-0006-8730-311X}{0009-0006-8730-311X}\\[1.2ex]
Aryeh Lev Zabokritskiy (Yohananov)\\
\normalsize Department of Computer Science\\
\normalsize Tel-Hai University of Kiryat Shmona in the Galilee\\
\normalsize MIGAL---Galilee Research Institute\\
\normalsize ORCID: \href{https://orcid.org/0000-0003-3151-6192}{0000-0003-3151-6192}}
\date{}

\begin{document}

\maketitle
\begin{abstract}
We study codes correcting up to \(t\geq1\) successive adjacent
transpositions, using block concatenation to support blockwise encoding and
decoding. A fixed set \(P\) of words of length \(b\) over a
\(q\)-symbol alphabet generates, for every \(m\), the length-\(mb\) code of all
concatenations of \(m\) words from \(P\). For our
explicit constructions, the prepared encoder and decoder depend only on the
block parameters and are reused at every concatenation length.
We first prove that correction at every concatenation length is certified by a single
finite test at \(2t+1\) blocks, and this test is tight over a \(q\)-symbol alphabet exactly when \(\binom q2\geq2t\).
For binary words and \(b\geq2t+1\), we give an
explicit construction with total redundancy at most
\(mt\log_2 b+O_t(m)\), matching the optimal per-block redundancy
\(t\log_2 b+O_t(1)\) among block sets correcting at every concatenation
length. Among all binary \(t\)-correcting codes of length \(mb\), the optimal total redundancy is
\(t\log_2(mb)+O_t(1)\). Using all concatenations of one block set makes our total
redundancy grow linearly in \(m\) and buys independent block encoding, with one encoder and
decoder at every length. The construction encodes and decodes in time linear in \(mb\) for
fixed \(b,t\), and recovers the \(j\)-th source block from the first \(jb\)
received symbols. We also prove that, whenever the length-\(mb\) code
corrects \(t\) transpositions, the \(j\)-th source block is determined
within \(2tb\) received symbols beyond position \(jb\), and that explicit binary
block sets with rates tending to one as \(b\) grows with \(t\) fixed attain this
bound in the worst case.
Finally, we prove that every ordinary correcting code contains a constant
fraction of its words whose arbitrary concatenations retain correction,
for fixed \(q,t\).

\end{abstract}


\section{Introduction}

Local reordering errors, which preserve the symbols of a word but may change
their order, arise in bit-shift channels~\cite{kovacevic2019runlength} and in
limited permutation channels~\cite{langberg2017lpc}. The corresponding coding
problems depend on which reorderings the channel permits. Successive
adjacent transpositions count each exchange of two neighboring symbols
separately and allow later exchanges to overlap earlier ones. Other models
restrict exchanges to disjoint pairs or bound the displacement of each
symbol~\cite{kovacevic2026bounds,langberg2017lpc}.

We study codes that correct up to \(t\geq1\) successive adjacent
transpositions and are built by block concatenation. We use a fixed
nonempty set \(P\subseteq\Sigma_q^b\) of blocks of length \(b\), where
\(\Sigma_q\) is a \(q\)-symbol alphabet, and write
\(\cC_{mb}(P)=P^m\) for all concatenations of \(m\) such blocks.
Each message block is encoded independently by the same map into a source block in \(P\).
Block concatenation permits the same prepared encoder and decoder to
serve every number of blocks. Our two explicit binary block constructions are the derivative and
endpoint families of Sections~\ref{sec:correcting-block-sets} and~\ref{sec:blockwise-decoding};
their preparation is deterministic and polynomial in \(b\) (searches over candidate defining
polynomials and field elements, fewer than \(2(b+1)\) each;
Section~\ref{sec:correcting-block-sets}), and the prepared
code data are reused when \(m\) changes. Each transmission is framed: the receiver's
initial side information consists of \(P\), \(b\), \(t\) and the total length \(n=mb\).
The composition of the transmitted word generally depends on the message.
The complete received word determines this composition because swaps preserve it,
whereas a proper received prefix may leave several compositions possible. The decoders of both families
process symbols as they arrive, using data memory bounded in terms of
\(b,t\), together with \(O(\log n)\) bits for framing counters.

We require correction of at most \(t\) successive adjacent
transpositions in the entire transmitted word, swaps across block
boundaries included; a swap of two equal symbols counts as an error but
leaves the word unchanged. The error budget \(t\) applies to the complete frame.
As in Kova\v{c}evi\'c's bounds~\cite{kovacevic2019runlength},
\(t\) is fixed while \(n=mb\) varies. Boundary swaps couple neighboring
blocks, so correction of isolated blocks alone is insufficient.
By the finite-locality theorem below, a single finite test, done once,
certifies correction at every concatenation length.

Adjacent transpositions are related to deletion and insertion errors:
the change \(ab\mapsto ba\) can be performed by deleting \(a\) and inserting
it after \(b\). Gabrys, Yaakobi, and Milenkovic combine a
Varshamov--Tenengolts decoding step with a transposition decoder to correct
one deletion and several adjacent transpositions~\cite{gabrys2018damerau}.
This combination is a further application of efficient transposition decoding,
beyond correcting transpositions alone.
For a single adjacent transposition,
Radonjic, Das, and Pokhrel give integer-code
constructions~\cite{radonjic2024integer}. Wu et al. study systematic
deletion-correcting codes for permutations and words of prescribed
composition~\cite{wu2026systematic}. Their fixed-composition setting also
provides block sets for the present channel
(Remark~\ref{rem:deletion-reduction}).

For binary words, the model of successive adjacent transpositions is the
bit-shift model studied by Kova\v{c}evi\'c~\cite{kovacevic2019runlength}: the error
budget is the sum of the absolute displacements of the ordered occurrences
of \(1\). His lattice constructions and packing bounds determine the
asymptotic order of the largest code for every fixed \(t\), giving optimal
redundancy \(t\log_2 n+O_t(1)\) for \emph{ordinary} binary codes of length \(n\) (codes
that are arbitrary subsets of \(\{0,1\}^n\); this sense of ``ordinary'' is
used throughout). We study the same channel with an additional requirement on how
codewords are assembled and decoded.

Block concatenation has been used for codewords whose blocks are permutations, with an
outer code~\cite{heymann2014concatenated}, and for the limited permutation
channel~\cite{langberg2017lpc,chee2016string}. Ben Shimon and Zabokritskiy
study binary and \(q\)-ary block-concatenation codes for the latter
channel~\cite{benshimon2026local,benshimon2026extinction}. Their channel
radius bounds the displacement of each symbol: already at radius one,
a received word may result from arbitrarily many disjoint adjacent swaps.
At radius one they give a local criterion that reduces correction at every length to finitely
many local checks between blocks~\cite{benshimon2026local}; they also give an exact finite
verifier of all-length correction for finite prefix-free binary block
sets~\cite[Section~6.2, Theorems~13--14]{benshimon2026local}.
Our finite test gives a uniform \(2t+1\)-block criterion for the different error set of at most
\(t\) successive adjacent swaps.
Kova\v{c}evi\'c, Kiah, and Goyal also give block constructions for the limited permutation
channel~\cite[Section~3.2]{kovacevic2026bounds}. Our codes share the
block-concatenation approach, but use a different error budget: two
disjoint swaps are two errors here, whereas both are permitted at radius
one in their model, and successive swaps here may also overlap.

We ask when
\(P\) corrects at every concatenation length, how large it can be, and
how soon each source block can be decoded.
We first prove that correction at \(2t+1\) blocks guarantees correction
at every concatenation length. The proof restricts a collision, two
distinct codewords that reach a common received word with at most \(t\)
swaps each, to a connected interval of the blocks touched by these swaps. The interval has at
most \(2t+1\) blocks because the two paths cross at most \(2t\) block boundaries; this count
also underlies the test's tightness and the delay bound.

The test is tight over a
\(q\)-symbol alphabet exactly when \(\binom q2\geq2t\): a block set
can then correct through \(2t\) blocks and first have a collision at
\(2t+1\), which needs a distinct unordered pair of symbols at each of its \(2t\) block
boundaries. Over smaller alphabets, binary alphabets included for every \(t\), a test at \(2t\)
blocks suffices; whether fewer blocks suffice is open when \(\binom q2\leq2t-2\), as for binary
words with \(t\geq2\) (Section~\ref{sec:open}). The tightness construction is a guarded
encoding of a labelled digraph
derived from the \v{C}ern\'y automaton~\cite{cerny1964}: each arc becomes a block whose
long constant internal factor of guard symbols forces colliding codewords to agree away
from block boundaries. The automaton has pairs of states whose shortest synchronizing words
take every length from one through \(\binom q2\)~\cite[pp.~2, 4]{behague2022synchronizing}
(Lemma~\ref{lem:cerny-pair-spectrum}). Two
walks that diverge into the pair with synchronization time \(2t\) give a first collision at
\(2t+1\) blocks. Its symbol pairs at the \(2t\) block boundaries are the distinct pairs of
states the walks visit, so sufficiency counts the same \(\binom q2\) pairs as necessity.

We give an explicit binary block construction for \(b\geq2t+1\), the
derivative family, with total redundancy at most
\(mt\lceil\log_2(b+1)\rceil\), compared with the optimum
\(t\log_2(mb)+O_t(1)\) for ordinary binary codes in
Kova\v{c}evi\'c's setting~\cite{kovacevic2019runlength}.
Its blocks are the derivatives (inverse prefix sums) of the codewords of a binary BCH
code~\cite{bose1960binary}. By the established prefix-sum
reduction~\cite{gabrys2018damerau,kovacevic2026bounds}, prefix sums of the whole received
word turn each swap, boundary swaps included, into at most one substitution in one block's
segment, so the \(j\)-th source block is recovered from the first \(jb\) received
symbols. Its correction at every concatenation length is thus proved directly, as for the endpoint
family. The finite test serves arbitrary block sets, which it certifies in principle by an
enumeration over \(P^{2t+1}\); its value is structural: it bounds the number of blocks that
must be examined.
Every binary block set correcting at every concatenation length is itself
a \(t\)-correcting code of length \(b\), so Kova\v{c}evi\'c's upper bound
shows that our per-block redundancy is optimal up to an additive
\(O_t(1)\) term; equivalently, the largest such block set has size
\(\Theta_t(2^b/b^t)\). Given this optimum, the gap in total redundancy is inherent to using
all concatenations of one block set: their redundancy is \(m\) times the per-block redundancy,
linear in \(m\), whereas the optimum for ordinary codes grows logarithmically. In exchange,
one block encoder and decoder serve every number of blocks. For an ordinary code, preparation,
encoding and decoding costs, and the position by which each block is
available depend on its construction (Table~\ref{tab:tradeoff}).

For the derivative family and the full transmitted word of length \(n=mb\), using schoolbook
arithmetic in a polynomial basis, the encoding and decoding costs after the reusable preparation
are \(O(nt\log b)\) and \(O\bigl((nt+mt^3)\log^2 b\bigr)\) bit operations,
respectively. For fixed \(t\), these processing costs are
\(O(n\log b)\) and \(O(n\log^2 b)\), as for the endpoint family (Table~\ref{tab:tradeoff});
they are linear in \(n\) when \(b\) is also fixed.

For decoding delay, we count the additional received symbols beyond
position \(jb\) needed to determine the \(j\)-th source block. Computation
steps are counted separately. Our delay results are stated for the receiver described above,
whose initial side information consists of \(P\), \(b\), \(t\) and \(n\).
We prove that whenever \(\cC_{mb}(P)\) is
\(t\)-correcting, the delay is at most \(2tb\), by the same count of block boundaries
crossed by the two paths. The derivative family has zero additional delay.
The endpoint family, at lengths \(b=2^s+1\), places the derivative of a BCH
codeword between two endpoint bits that let an uncertainty persist across successive
blocks. Its worst-case delay is exactly \(2tb\) in frames of at least \(2t+1\) blocks, so
the general delay bound is sharp even at rates tending to one; in frames of at least \(2t+2\)
blocks, it is \(2tb\) even for a receiver told the composition
(Section~\ref{sec:blockwise-decoding}).

At these block lengths, Proposition~\ref{prop:constituent-code-transfer}, applied to the endpoint family's BCH code with
two zero coordinates appended, gives a block set of the same length and size with zero
additional delay. The derivative family of Table~\ref{tab:tradeoff} instead uses a longer,
shortened BCH code and has bound \(t(s+1)\) at these block lengths. The endpoint family's
guarantee \(st+2\), smaller when \(t\geq3\), thus reflects the choice of BCH constituent.
Its role here is to attain the general delay bound.

Finally, we prove that, for general alphabets, every ordinary
\(t\)-correcting code \(\cD\subseteq\Sigma_q^b\) contains a subset
\(P\) of size at least \(c_{q,t}|\cD|\) whose concatenations correct
at every length, where \(c_{q,t}=\mu^{-(q-1)}\) with \(\mu=\lfloor\sqrt{2t}\rfloor+1\).
To select \(P\), we group the codewords of \(\cD\) by their symbol counts modulo \(\mu\) and keep
a largest group, whose congruent compositions extend its correction at length \(b\) to every length.
If \(h\) is half the \(\ell_1\)-distance between two words' prefix symbol-count vectors at a cut,
then the words are at least \(h^2\) swaps apart (Lemma~\ref{lem:prefix-flow-amplitude}).
Consequently, two source concatenations in \(P^m\) at distance at most \(2t\) have equal prefix
compositions at every block boundary: otherwise congruence modulo \(\mu\) gives \(h\geq\mu\),
forcing their distance to be at least \(\mu^2>2t\).
Thus, relative to \(\cD\), arbitrary concatenation costs only a constant additional
redundancy per block for fixed \(q,t\). The guarantee covers correction and size;
efficient selection and encoding of the subset depend on the
representation of the supplied code.

Our contributions are as follows.
\begin{enumerate}[label=(\roman*)]
\item We prove that correction at \(2t+1\) blocks guarantees correction
at every length, and that this finite test is tight precisely when
\(\binom q2\geq2t\), with explicit block sets attaining it.
\item We construct binary codes of length \(mb\) with total redundancy
at most \(mt\lceil\log_2(b+1)\rceil\) for \(b\geq2t+1\), and give
encoding and decoding algorithms that run in time linear in \(mb\) for
fixed \(b,t\) and decode the \(j\)-th source block from the first \(jb\)
received symbols.
\item We prove that whenever \(\cC_{mb}(P)\) is \(t\)-correcting, the
\(j\)-th source block is determined after at most \(2tb\) additional received symbols
beyond position \(jb\), and construct an explicit binary family, the endpoint family,
whose worst-case delay equals this bound, with block redundancy \(t\log_2 b+O_t(1)\).
\item We prove that every ordinary correcting code \(\cD\) contains a
subset \(P\) with \(|P|\geq c_{q,t}|\cD|\) whose concatenations correct
at every length, where \(c_{q,t}>0\) is independent of \(b\).
\end{enumerate}

Table~\ref{tab:tradeoff} summarizes these redundancy and processing guarantees.

\begin{table}[htbp]
\centering\small
\caption{Ordinary binary codes of optimal
redundancy~\cite{kovacevic2019runlength} versus the two explicit binary
block families of this paper, for fixed \(t\) and total length
\(n=mb\): the derivative family and the endpoint family. The derivative family requires
\(b\geq2t+1\). The endpoint family uses \(b=2^s+1\), with
\(2t<2^s-1\) and \(2^s-1-st\geq2\); there its smaller redundancy bound for \(t\geq3\) comes from
its choice of BCH code (Section~\ref{sec:blockwise-decoding}).
Processing costs count bit operations under schoolbook arithmetic; per-block quantities refer
to the indicated block families. Block \(j\) is the \(j\)-th segment of \(b\) symbols of the transmitted word.}
\label{tab:tradeoff}
\begin{tabular*}{\columnwidth}{@{\extracolsep{\fill}}llll}
\toprule
 & Ordinary code & Derivative family & Endpoint family \\
\midrule
Total redundancy & \(t\log_2 n+O_t(1)\) &
 \(\leq mt\lceil\log_2(b{+}1)\rceil\) &
 \(\leq m\bigl(t\log_2(b{-}1)+2\bigr)\) \\
Per-block redundancy & --- &
 \(\leq t\lceil\log_2(b{+}1)\rceil\) &
 \(\leq t\log_2(b{-}1)+2\) \\
Per-block optimum & --- & met within \(O_t(1)\) & met within \(O_t(1)\) \\
Preparation & code-dependent & once, for \((b,t)\) & once, for \((b,t)\) \\
Encoding (fixed \(t\)) & code-dependent & \(O(n\log b)\) & \(O(n\log b)\) \\
Decoding (fixed \(t\)) & code-dependent & \(O(n\log^2 b)\) & \(O(n\log^2 b)\) \\
Block \(j\) available by & code-dependent & symbol \(jb\) &
 \(\min\{n,(j{+}2t)b\}\) \\
Correction for every \(m\) & not required & yes & yes \\
\bottomrule
\end{tabular*}
\end{table}

\FloatBarrier
Section~\ref{sec:model} defines the channel and proves finite locality.
Section~\ref{sec:guarded-labelled-automata} proves tightness of the finite
test and determines the alphabet threshold.
Section~\ref{sec:correcting-block-sets} gives the explicit binary
construction and its encoding and decoding algorithms.
Section~\ref{sec:blockwise-decoding} studies decoding delay.
Section~\ref{sec:extraction} extracts block sets correcting at every
concatenation length from ordinary codes and compares their sizes.
Section~\ref{sec:open} concludes with the remaining questions.

\section{Channel model and finite locality}
\label{sec:model}

We formulate correction in terms of the possible predecessors of a received
word, then show that every collision among concatenations of blocks from a
block set can be restricted
to a collision on a bounded interval of blocks.  This gives a finite test whose number of blocks
is bounded independently of the block length.

\subsection{Channel model and first collision powers}

Let \(\Sigma\) be a finite alphabet, and let
\(x=x_1\cdots x_n\in\Sigma^n\).  For \(1\leq i<n\), let
\(\tau_i(x)\) be the word obtained from \(x\) by swapping \(x_i\) and \(x_{i+1}\).  The
adjacent-transposition graph has vertex set \(\Sigma^n\) and an edge
between two distinct words when one is obtained from the other by one
adjacent swap. A swap of two equal symbols counts as a channel error but
leaves the word unchanged, so the graph has no edge for it; the paragraph
after the next definition justifies this omission.
Write \(\AT(x,z)\) for its shortest-path distance, with
\(\AT(x,z)=\infty\) if \(x\) and \(z\) lie in different components,
equivalently if some symbol has different multiplicities in the two words.
This adjacent-transposition metric is also known as the Kendall tau
sequence distance~\cite{cicirello2020kendall}.
For any word \(w\), write
\(\operatorname{comp}(w)\in\mathbb N_0^{\Sigma}\) for its composition
vector, whose coordinate indexed by \(a\in\Sigma\) counts the occurrences
of \(a\) in \(w\).  Let \(e_a\) denote the corresponding standard basis
vector.

\begin{definition}[Predecessor balls and correction]
For an integer \(t\geq0\) and a received word \(y\in\Sigma^n\), define
\[
B_t^{\mathrm{AT}}(y)=
\left\{x\in\Sigma^n:\,
y=\tau_{i_s}\cdots\tau_{i_1}(x)
\text{ for some }0\leq s\leq t
\text{ and }i_1,\ldots,i_s\in\{1,\ldots,n-1\}
\right\}.
\]
For \(s=0\), the product of transpositions is the identity.  A code
\(\cD\subseteq\Sigma^n\) \emph{corrects \(t\) adjacent transpositions},
or is \emph{\(t\)-correcting}, if
\[
|\cD\cap B_t^{\mathrm{AT}}(y)|\leq1
\qquad\text{for every }y\in\Sigma^n.
\]
Thus \(B_t^{\mathrm{AT}}(y)\) contains the words that can reach \(y\)
using at most \(t\) swaps. A code is \(t\)-correcting when every received
word has at most one codeword predecessor in this ball.
\end{definition}

For block concatenations, fix a nonempty set \(P\subseteq\Sigma^b\),
where \(b\geq1\). Write
\(\cC_{mb}(P)=P^m\) for the set of all concatenations of \(m\geq1\)
blocks from \(P\). Each codeword has total length \(n=mb\), and
\(|\cC_{mb}(P)|=|P|^m\). Thus \(m\) is the number of concatenated
blocks and \(|P|\) is the number of distinct blocks available at each
position. Because every block has the same length \(b\), each word in \(P^m\) has a
unique decomposition into \(m\) blocks. For \(x=x^{(1)}\cdots x^{(m)}\)
and \(z=z^{(1)}\cdots z^{(m)}\) in \(P^m\), the blocks
\(x^{(j)}\) and \(z^{(j)}\) correspond to the same block position.
The \(i\)-th block boundary, \(1\le i<m\), lies between blocks
\(i\) and \(i+1\); the adjacent coordinates are positions \(ib\) and \(ib+1\).
For \(1\le j<n\), cut~\(j\) separates positions \(j\) and \(j+1\), and the
block boundaries are the fixed cuts \(b,2b,\ldots,(m-1)b\).
The channel acts on all words in \(\Sigma^{mb}\); intermediate and received words
need not belong to \(P^m\).
Reversibility lets us view the words reachable from \(y\) in at most \(t\)
swaps as its possible predecessors; decoding asks which of them is a codeword.  The decoder knows the
block set \(P\), its block length \(b\), and the error radius \(t\); the
received length determines the power \(m\).  Each transposition counts as one
error.  In particular, two disjoint
swaps count as two errors.  A swap sequence may repeat indices and may
contain overlapping swaps. A transposition of two equal adjacent symbols counts as an error but
leaves the current word unchanged; deleting such steps preserves every
subsequent state and the final output, so omitting them from the graph
does not change which words can be reached with at most \(t\) swaps.
Consequently,
\[
B_t^{\mathrm{AT}}(y)=\{x\in\Sigma^n:\AT(x,y)\leq t\}.
\]
A language \(L\subseteq\Sigma^*\) is \(t\)-correcting
when the code \(L\cap\Sigma^n\) is \(t\)-correcting for every length
\(n\).  Let \(P^*=\bigcup_{m\geq0}P^m\). Then
\(P^*\cap\Sigma^{mb}=P^m\), and every word in \(P^*\) has length divisible by \(b\). In particular, \(P^*\) is \(t\)-correcting if and only if \(P^m\) is
\(t\)-correcting for every \(m\geq1\); we write \(P^*\) for statements
that hold uniformly over all concatenation lengths, and \(P^m\)
(equivalently \(\cC_{mb}(P)\)) when the length is fixed.

\begin{lemma}[Correction and minimum distance]
\label{lem:distance}
A code \(\cD\subseteq\Sigma^n\) is \(t\)-correcting if and only if
\[
\AT(x,z)\geq 2t+1
\qquad\text{for all distinct }x,z\in\cD.
\]
In particular, correcting two successive adjacent transpositions is equivalent to
minimum adjacent-transposition distance at least five.
\end{lemma}

\begin{proof}
If \(x,z\in B_t^{\mathrm{AT}}(y)\), the two paths through \(y\) give
\(\AT(x,z)\leq 2t\).  Conversely, let a shortest path from \(x\) to \(z\)
have length at most \(2t\).  Choose a vertex \(y\) on this path after at
most \(t\) edges and with at most \(t\) edges remaining.  Then
\(x,z\in B_t^{\mathrm{AT}}(y)\).
\end{proof}

A \emph{radius-\(t\) collision} in a set \(\cD\subseteq\Sigma^n\)
is a pair of distinct words \(x,z\in\cD\) together with a word
\(y\in\Sigma^n\) such that \(x,z\in B_t^{\mathrm{AT}}(y)\), written
\(x\to y\leftarrow z\). Such a received word \(y\) exists if and only if \(\AT(x,z)\leq2t\),
by the two directions in the proof of Lemma~\ref{lem:distance}.

Fix integers \(t\geq1\) and \(q\geq2\), write
\(\Sigma_q=\{0,1,\ldots,q-1\}\), and let
\(P\subseteq\Sigma_q^b\) be nonempty.  Define its \emph{first collision power}
at radius \(t\) by
\[
m_t(P)=\min\{m\geq1:P^m\text{ contains a radius-\(t\) collision}\}.
\]
When every power is \(t\)-correcting, set \(m_t(P)=\infty\). For each \(t\geq1\), the derivative family
of Section~\ref{sec:correcting-block-sets} contains explicit binary block sets with
\(m_t(P)=\infty\) and arbitrarily large \(|P|\), the number of distinct blocks.
If a collision occurs, appending the same block to both predecessors and
the received word preserves it, so collisions persist in every larger power.
Let \(D_t(q)\) be the supremum of the finite first collision power over all block lengths and block sets:
\begin{equation}
\label{eq:def-depth}
D_t(q)=
\sup_{\substack{b\geq1,\ \varnothing\neq P\subseteq\Sigma_q^b\\
                 m_t(P)<\infty}}
m_t(P).
\end{equation}
The indexing family is nonempty: for distinct \(\alpha,\beta\in\Sigma_q\), the
block set \(P=\{\alpha,\beta\}\subseteq\Sigma_q^1\) has \(m_t(P)=2\).  Indeed,
its first power is \(t\)-correcting, while \(\alpha\beta\) and \(\beta\alpha\) differ by one adjacent
transposition.
For the finite test, once Theorem~\ref{thm:five} shows that \(D_t(q)\) is finite, it is
the number of blocks that a test must cover to certify every block set over
\(\Sigma_q\); the discussion after that theorem makes this precise.
\subsection{Finite locality}
\label{sec:general}

The channel preserves length, so a collision in \(P^*\) always compares
two words in one power \(P^m\). The next theorem shows that correction of
one power certifies correction at every concatenation length. The proof uses a graph on
block positions, with each swap joining the blocks it touches: a component
containing a differing block carries a collision on at most one more block
than its number of swaps.

\begin{theorem}[Finite locality]
\label{thm:five}
Let \(P\subseteq\Sigma^b\) be nonempty and let \(t\geq1\) be an integer.
With at most \(t\) successive adjacent transpositions on each predecessor path,
\[
P^*\text{ is \(t\)-correcting}\quad\Longleftrightarrow\quad
P^{\,2t+1}\text{ is \(t\)-correcting}.
\]
In particular, \(P^*\) corrects two successive adjacent transpositions
if and only if \(P^5\) does.
\end{theorem}

\begin{proof}
Suppose that distinct \(x,z\in P^m\) reach one received word \(y\), using
at most \(t\) swaps on each path.  Write
\(x=x^{(1)}\cdots x^{(m)}\) and
\(z=z^{(1)}\cdots z^{(m)}\), with all displayed factors in \(P\).
For \(1\leq j\leq m\), let \(I_j=\{(j-1)b+1,\ldots,jb\}\) be the
\(j\)-th block interval.  Thus \(x^{(j)}\) and \(z^{(j)}\) occupy
the coordinates of \(I_j\).  Form a multigraph \(G\) on \(\{1,\ldots,m\}\),
with one edge occurrence for each swap used by either path.
A swap inside block \(j\) gives a loop at \(j\); a swap across the boundary
between blocks \(j\) and \(j+1\) gives an edge \(\{j,j+1\}\).
Since each path uses at most \(t\) swaps, \(G\) contains at most \(2t\)
edge occurrences.

On every block untouched by the swaps, both paths preserve the restricted word, so \(x,z,y\) agree there.  Choose \(j\) with
\(x^{(j)}\neq z^{(j)}\); at least one used swap touches block \(j\).
Let \(K\) be the vertex set of its connected component in \(G\).
Edges between distinct vertices join
consecutive blocks, so \(K\) is a contiguous interval.  Write \(K=\{r_0,r_0+1,\ldots,s_0\}\), put
 \(I=\bigcup_{j\in K}I_j\), and define
 \(x_K=x^{(r_0)}\cdots x^{(s_0)}\),
 \(z_K=z^{(r_0)}\cdots z^{(s_0)}\), and \(y_K=y|_I\), with the
inherited coordinate order.  Every used swap is supported wholly inside \(I\)
or wholly outside it, since a swap crossing an endpoint of \(I\) would join
\(K\) to another vertex.  Projecting each intermediate word onto \(I\) turns
an outside swap into an unchanged step and an inside swap into the corresponding
adjacent transposition on the restricted word.  Deleting the unchanged steps from
the two projected paths therefore gives paths of adjacent swaps
\[
x_K\longrightarrow y_K\longleftarrow z_K,
\qquad x_K\neq z_K.
\]
If the component contains \(e\) edge occurrences, it has at most \(e+1\)
vertices, since a spanning tree uses \(|K|-1\) of those edge occurrences.  Hence \(K\) spans at most
\(2t+1\) blocks.

Thus a collision in \(P^*\) yields a collision in some \(P^k\), where
 \(k=|K|\leq 2t+1\).  Choose any \(p\in P\).  Appending \(2t+1-k\)
 identical copies of \(p\) to both predecessors and the received word pads this collision to
\(P^{2t+1}\).  The reverse implication is immediate.
\end{proof}

In terms of \eqref{eq:def-depth}, Theorem~\ref{thm:five} gives
\[
D_t(q)\leq2t+1
\qquad(t\geq1,\ q\geq2).
\]
Because the indexing family in \eqref{eq:def-depth} is nonempty and its
finite integer values now lie in \(\{1,\ldots,2t+1\}\), the supremum is
attained; that is, \(D_t(q)\) is the largest finite first collision power.
Thus \(D_t(q)=d\) means that testing through power \(d\) always
suffices and that some block set has its first collision exactly in power
\(d\); whether \(d<2t+1\) depends on \(q\), as
Section~\ref{sec:guarded-labelled-automata} shows.

Operationally, Theorem~\ref{thm:five} gives a finite test for correction at
every concatenation length: it suffices to test the single
power \(P^{2t+1}\). A direct testing procedure follows by enumeration.
Put \(m=2t+1\), and \(n=mb\). Generate every word of \(P^m\)
and every sequence of at most \(t\) adjacent swap indices, keeping each
resulting word together with its predecessor. There are at most
\[
 |P|^{2t+1}\sum_{s=0}^{t}(n-1)^s
\]
such pairs. Grouping pairs with the same resulting word determines whether any
received word has two distinct predecessors.

To determine when the finite test is tight, we will use the following
cut estimate. It lower-bounds the number of swaps by the composition
imbalances across prefixes. For a vector \(v\) indexed by \(\Sigma\),
write \(\lVert v\rVert_1=\sum_{\sigma\in\Sigma}|v_\sigma|\).

\begin{lemma}[A lower bound from prefix compositions]
\label{lem:prefix-flow}
Let \(x,y\in\Sigma^n\) have the same composition, and for
\(1\leq j<n\) put
\[
 \Delta_j=
 \operatorname{comp}(x_1\cdots x_j)
 -\operatorname{comp}(y_1\cdots y_j).
\]
Then
\begin{equation}
 \AT(x,y)
 \geq \sum_{j=1}^{n-1}\frac12\lVert\Delta_j\rVert_1.
 \label{eq:prefix-flow}
\end{equation}
\end{lemma}

\begin{proof}
An adjacent transposition across cut \(j\), the boundary between positions
\(j\) and \(j+1\), changes only the prefix-composition difference at that
cut, by a vector \(\pm(e_\alpha-e_\beta)\) of
\(\ell_1\)-norm at most two.
Let \(x=w^{(0)},w^{(1)},\ldots,w^{(L)}=y\) be a shortest
path, and let \(N_j\) be the number of its swaps crossing cut \(j\).
At \(w^{(0)}=x\) the prefix difference is \(\Delta_j\); at \(w^{(L)}=y\) every
prefix of the word equals the corresponding prefix of \(y\), so the prefix difference is \(0\).  Hence \(\lVert\Delta_j\rVert_1\le 2N_j\), that is
\(N_j\geq\lVert\Delta_j\rVert_1/2\).  Every swap crosses exactly one
cut, and \(L=\AT(x,y)\), so
summing over \(j\) proves \eqref{eq:prefix-flow}.
\end{proof}

\section{Tightness and alphabet size}
\label{sec:guarded-labelled-automata}
\label{sec:tq-bounds}

The next theorem characterizes when \(D_t(q)\) from \eqref{eq:def-depth}
attains \(2t+1\). A first collision at that power uses one swap at each
of the \(2t\) block boundaries, which is the whole swap budget of the two
predecessor paths, and the unordered pairs of symbols exchanged at these
boundaries must be distinct: repeating one in either orientation would give a
collision in a smaller power. Whatever the block length, there are only
\(\binom q2\) unordered symbol pairs, so longer blocks cannot supply the \(2t\)
distinct pairs this needs.
Put
\[
 \kappa_t=\min\left\{q\geq2:\binom q2\geq2t\right\}
 =\left\lceil\frac{1+\sqrt{1+16t}}2\right\rceil.
\]

\begin{theorem}[Exact alphabet threshold]
\label{thm:exact-alphabet-threshold}
For every \(t\geq1\) and \(q\geq2\),
\[
 D_t(q)=2t+1\quad\Longleftrightarrow\quad\binom q2\geq2t.
\]
Consequently, the smallest alphabet size attaining the locality bound is
\[
 Q_t:=\min\{q\geq2:D_t(q)=2t+1\}=\kappa_t.
\]
\end{theorem}

For sufficiency, we encode labelled transitions as blocks so that two
walks with the same endpoints and label word yield a channel collision.
Adding two equally labelled start arcs to the \v{C}ern\'y automaton
lets the walks diverge, after which a state pair synchronizes in the
required number of steps. We give this construction first and then prove
the alphabet obstruction.
This section also determines \(D_t(q)\) immediately below the
threshold, when \(\binom q2=2t-1\).

\subsection{Guarded encoding}

Encode a directed arc by its tail, a long constant internal factor representing its
label, and its head. For \(x,z\in P^m\), the symbols at the \(i\)-th block boundary are reversed between \(x\) and \(z\) when
\(x_{ib}\ne x_{ib+1}\) and
\((z_{ib},z_{ib+1})=(x_{ib+1},x_{ib})\). Long constant factors force two nearby concatenations to have
the same aligned labels and to differ only by reversals of the symbols at block boundaries.
On a run of consecutive blocks whose internal boundary pairs are all reversed, taking the
arcs alternately from the two concatenations, starting at the run's first block with either
one, gives two directed walks.

Let \(\mathcal A=(V,E,\lambda)\) be a finite labelled directed
multigraph, with maps \(\operatorname{tail},\operatorname{head}:E\to V\)
and \(\lambda:E\to\Lambda\), such that
\begin{equation}
 e\longmapsto\bigl(\operatorname{tail}(e),\lambda(e),
                         \operatorname{head}(e)\bigr)
 \quad\text{is injective}.
 \label{eq:no-label-parallel-arcs}
\end{equation}
Assume \(V\subseteq\Sigma\) and choose distinct symbols
\(g_\ell\in\Sigma\), one for each \(\ell\in\Lambda\). These guard
symbols may also be vertices, since the estimates below allow arbitrary
prefix-composition imbalance at the start of an internal factor.
For \(L\geq1\), define
\[
 \operatorname{enc}_L(e)=\operatorname{tail}(e)\,
 g_{\lambda(e)}^L\,\operatorname{head}(e),\qquad
 P(\mathcal A,L)=\{\operatorname{enc}_L(e):e\in E\}
 \subseteq\Sigma^{L+2}.
\]
Write \(\operatorname{amb}(\mathcal A)\) for the minimum length of two
directed walks with different arc sequences and the same initial vertex,
final vertex, and label word, with the common endpoints ranging over \(V\) and possibly
equal, or \(\infty\) if there are no such walks.
The distinct guard symbols and \eqref{eq:no-label-parallel-arcs} ensure
that different arcs have different encoded blocks.
For two concatenations of encoded blocks at distance at most \(2t\), the first
of the sufficient length conditions below rules out that a nonzero prefix-composition imbalance
between them persists across a common internal factor, and the second rules
out that their aligned internal labels differ.

\begin{theorem}[Guarded encoding of labelled digraphs]
\label{thm:guarded-labelled-arc-transfer}
Let \(t\geq1\), put \(M=2t\), and suppose that \(E\neq\varnothing\)
and
\begin{equation}
 L+1>M,\qquad
 \left\lfloor\frac{(L+1)^2}{4}\right\rfloor>M.
 \label{eq:guard-length-conditions}
\end{equation}
Then
\begin{equation}
 m_t\bigl(P(\mathcal A,L)\bigr)=
 \begin{cases}
  \operatorname{amb}(\mathcal A),&\operatorname{amb}(\mathcal A)\leq M+1,\\
  \infty,&\operatorname{amb}(\mathcal A)>M+1.
 \end{cases}
 \label{eq:exact-guarded-transfer}
\end{equation}
Every pair of distinct directed walks of length at most \(M+1\) with the
same initial vertex, final vertex, and label word gives two
predecessor paths with at most \(t\) swaps each to a common received word.
\end{theorem}

\begin{proof}
Suppose that distinct \(x,z\in P(\mathcal A,L)^r\) reach the same
received word with at most \(t\) swaps each. Then
\(\AT(x,z)\leq M\), and \(x,z\) have the same composition.
If the internal factor at an aligned position is \(\alpha^L\) in \(x\) and \(\beta^L\) in
\(z\), with \(\alpha\ne\beta\), the prefix differences at the
\(L+1\) cuts through this factor are
\[
 \Delta^{(s)}=\Delta^{(0)}+s(e_\alpha-e_\beta),\qquad 0\leq s\leq L.
\]
With \(c=\Delta^{(0)}_\alpha-\Delta^{(0)}_\beta\),
Lemma~\ref{lem:prefix-flow} gives
\[
 \AT(x,z)\geq\frac12\sum_{s=0}^L|c+2s|
 \geq\min_{\xi\in\mathbb R}\sum_{s=0}^L|s-\xi|
 =\left\lfloor\frac{(L+1)^2}{4}\right\rfloor>M.
\]
Thus the internal factors of corresponding blocks agree.

The remaining coordinates form clusters: the initial tail, the
head--tail symbol pair at each block boundary, and the final head. If the
first cluster with different multisets precedes the final head, the earlier
clusters cancel, leaving a nonzero integer prefix difference of coordinate
sum zero and \(\ell_1\)-norm at least two; this difference persists
across all \(L+1\) cuts of the following common internal factor, contributing
at least \(L+1>M\) to \eqref{eq:prefix-flow}. Hence all these clusters
have equal multisets; equal total composition then gives equality at the
final head, which has no following internal factor. The outer endpoints therefore agree, and the symbol pairs at
each block boundary are either identical or reversed with unequal symbols.
If \(c_0\) boundary pairs are reversed, their disjoint swaps transform
\(x\) into \(z\). Each such boundary contributes one to the sum in \eqref{eq:prefix-flow} and every other cut
contributes zero, whence
\begin{equation}
 \AT(x,z)=c_0\leq M.
 \label{eq:exact-seam-distance}
\end{equation}

Split the two block sequences at every boundary where the symbol pairs agree
coordinatewise, and select one of the resulting segments (a maximal interval of consecutive blocks,
bounded by such boundaries or by the endpoints of the word) containing a reversed boundary pair.
It has \(a\geq2\) blocks, and the symbol pairs at all \(a-1\) internal boundaries are reversed. Write
\(\varepsilon_1,\ldots,\varepsilon_a\) and
\(\varphi_1,\ldots,\varphi_a\) for the arcs encoded by its blocks
in \(x\) and \(z\). The labels agree at each position, the initial
tails and final heads agree by the equalities at the cuts or word endpoints,
and the reversed boundary pairs give
\[
 \operatorname{head}(\varepsilon_i)=\operatorname{tail}(\varphi_{i+1}),
 \qquad
 \operatorname{head}(\varphi_i)=\operatorname{tail}(\varepsilon_{i+1})
 \quad(1\leq i<a).
\]
Thus the alternating arc sequences
\[
 \varepsilon_1,\varphi_2,\varepsilon_3,\ldots
 \quad\text{and}\quad
 \varphi_1,\varepsilon_2,\varphi_3,\ldots
\]
are directed walks with the same label word and endpoints, since their
initial tails agree and either possible final arc has the common final head.
Equality of these walks would give \(\varepsilon_i=\varphi_i\) at every
position, leaving no reversed boundary pair, so they are distinct. By \eqref{eq:exact-seam-distance},
\(a-1\leq c_0\leq M\), so every collision in power \(r\) implies
\begin{equation}
 \operatorname{amb}(\mathcal A)\leq\min\{r,M+1\}.
 \label{eq:collision-implies-ambiguity}
\end{equation}

Conversely, take distinct same-label walks
\[
 p_0\xrightarrow{\varepsilon_1}p_1\xrightarrow{\varepsilon_2}\cdots
 \xrightarrow{\varepsilon_a}p_a,\qquad
 q_0\xrightarrow{\varphi_1}q_1\xrightarrow{\varphi_2}\cdots
 \xrightarrow{\varphi_a}q_a,
\]
where \(p_0=q_0\), \(p_a=q_a\), and \(a\leq M+1\). Form \(x\)
from the encoded \(\varepsilon_i\) at odd positions and encoded
\(\varphi_i\) at even positions, and form \(z\) by the opposite
choice. Injectivity in \eqref{eq:no-label-parallel-arcs} makes \(x\ne z\).
At block boundary \(i\), the words have \(p_i\mid q_i\) and
\(q_i\mid p_i\), in parity-dependent order, and they agree elsewhere.
Swap the symbols at every odd boundary with \(p_i\ne q_i\) in \(x\), and at every even such
boundary in \(z\). The resulting words agree coordinate by coordinate, and
each path uses at most
\[
 \left\lceil\frac{a-1}{2}\right\rceil\leq t
\]
adjacent transpositions of unequal symbols. This constructs a common received
word; together with \eqref{eq:collision-implies-ambiguity}, it proves
\eqref{eq:exact-guarded-transfer}.
\end{proof}

\subsection{Construction from the \v{C}ern\'y automaton}

To obtain two distinct walks of length \(R+1\) with the same endpoints and
label word, we make them diverge from a common state in one step, then
synchronize their two states in exactly \(R\) further steps. The \v{C}ern\'y automaton provides pairs
with every synchronization time from one through \(\binom q2\).
Under guarded encoding, the unordered symbol pair at the \(i\)-th block boundary of the
collision built from the two walks is the pair of states they occupy after \(i\) steps.
Along a shortest synchronizing word these pairs are pairwise distinct, since a repeat would
shorten the word. For \(R=2t\), the construction thus exhibits \(2t\) distinct pairs among the
\(\binom q2\), as Lemma~\ref{lem:maximal-witness} requires of every first collision at
\(2t+1\) blocks, and both directions of Theorem~\ref{thm:exact-alphabet-threshold} count
the same \(\binom q2\) pairs.
Fix \(q\geq2\), identify \(\Sigma_q\) with
\(\mathcal Q_q=\mathbb Z/q\mathbb Z\),
and use the transformations~\cite{cerny1964}
\begin{equation}
 a(i)=i+1\pmod q,\qquad c(q-1)=0,\qquad c(i)=i\quad(0\leq i<q-1).
 \label{eq:cerny-transformations}
\end{equation}
In words over \(\{a,c\}\), transformations act from left to right.
For a two-element set \(U\subseteq\mathcal Q_q\), let \(R(U)\) be the
minimum length of a word in \(\{a,c\}\) mapping \(U\) to a singleton.

\begin{lemma}[\v{C}ern\'y pairs]
\label{lem:cerny-pair-spectrum}
The values \(R(U)\) are \(1,2,\ldots,\binom q2\), each once.
If \(q=2r+1\), every pair has a unique representation
\begin{equation}
 U_{i,s}=\{i,i+s\pmod q\},\qquad
 0\leq i<q,\quad1\leq s\leq r,\qquad R(U_{i,s})=sq-i.
 \label{eq:cerny-nondiameter-time}
\end{equation}
If \(q=2r\), the same formula holds for \(1\leq s<r\), and the
diameters \(H_i=\{i,i+r\}\), \(0\leq i<r\), satisfy
\begin{equation}
 R(H_i)=\binom q2-i.
 \label{eq:cerny-diameter-time}
\end{equation}
The representation chooses the shorter cyclic direction; for diameters the
two directions tie, so \(0\leq i<r\) selects one representative.
\end{lemma}

Behague and Johnson show that in every deterministic automaton on
\(q\) states that admits a word mapping all states to a singleton,
the minimum word length mapping any prescribed pair of distinct states
to a singleton is at most \(\binom q2\), and that this bound on pair
rendezvous time is attained by a pair in the \v{C}ern\'y automaton
\cite[pp.~2, 4]{behague2022synchronizing}. Along a shortest word mapping a pair to
a singleton, the images of the pair before the last letter are distinct pairs.
For a pair attaining this bound they are therefore all \(\binom q2\) pairs, and the
suffixes of such a word, which are shortest words for these images, give each time
in the stated range exactly once. For odd \(q\), Gusev describes the pairs and the
action of the two letters on them, in a presentation that agrees with ours after
shifting state indices and interchanging the letters
\cite[Lemma~1, arXiv pp.~5--6]{gusev2014random}. We derive the synchronization times \(R(U)\), including those of even diameter pairs, and shortest
words from the transformations above alone; Gusev's description serves only to
match his presentation with ours.
We use the derived formulas and shortest words to choose a pair with a prescribed
synchronization time and to count the transitions retained in the construction of
Theorem~\ref{thm:cerny-guarded-construction}.

\begin{proof}
Let \(F\) equal the claimed value on each pair and zero on singletons.
We check that
\begin{equation}
 F(fU)\geq F(U)-1\qquad(f\in\{a,c\}).
 \label{eq:cerny-potential-descent}
\end{equation}
For nondiameters, \(aU_{i,s}=U_{i+1,s}\) decreases \(F\) by one
unless \(i=q-1\), when wrapping to \(U_{0,s}\) increases it. The map
\(c\) fixes pairs avoiding \(q-1\). It sends
\(U_{q-1,s}=\{q-1,s-1\}\) to \(U_{0,s-1}\), with
\(U_{0,0}\) denoting a singleton, and decreases \(F\) by one.
When \(q-1\) is the second endpoint, the image of \(U_{q-1-s,s}\)
under \(c\) is \(\{0,q-1-s\}\). Its canonical representation is
\(U_{q-1-s,s+1}\), except when \(q=2r+1\) and \(s=r\), giving
\(U_{0,r}\), or when \(q=2r\) and \(s=r-1\), giving \(H_0\).
Substitution in the corresponding formulas for \(F\) shows that
\(F\) does not decrease.
For even diameters, \(aH_i=H_{i+1}\) decreases \(F\) by one for
\(i<r-1\); the wrap to \(H_0\) does not decrease it. The map \(c\)
fixes \(H_i\) for \(i<r-1\), and sends \(H_{r-1}\) to
\(U_{0,r-1}\), or to a singleton when \(r=1\), decreasing \(F\)
by one. This proves \eqref{eq:cerny-potential-descent}, and hence
\(R(U)\geq F(U)\).

The words
\begin{equation}
 a^{q-1-i}c\,(a^{q-1}c)^{s-1}
 \label{eq:cerny-nondiameter-descent}
\end{equation}
and
\begin{equation}
 a^{r-1-i}c\,(a^{q-1}c)^{r-1}
 \label{eq:cerny-diameter-descent}
\end{equation}
synchronize \(U_{i,s}\) and \(H_i\), respectively, in the claimed
number of steps. The first factor sends the pair to \(U_{0,s-1}\)
or \(U_{0,r-1}\); each remaining factor sends
\(U_{0,j}\) to \(U_{0,j-1}\). Thus \(R=F\).
Each nondiameter layer \(s\) gives the interval
\((s-1)q+1,\ldots,sq\); for even \(q\), the diameter layer supplies
the remaining values through \(\binom q2\). These disjoint intervals
give the asserted spectrum, including \(q=2\).
\end{proof}

We retain a set of transitions containing those used by the chosen pair's shortest word, and add two
equally labelled start arcs. The resulting pair of walks gives the
desired first collision power under guarded encoding, whose distinct guard
symbols in \(\Sigma_q\) for the three labels require \(q\geq3\).

\begin{theorem}[Construction from the \v{C}ern\'y automaton]
\label{thm:cerny-guarded-construction}
Let \(t\geq1\), \(q\geq3\), and
\(1\leq R\leq\min\{2t,\binom q2\}\). There is a block set
\(P_{t,q,R}\subseteq\Sigma_q^{b_t}\) with
\begin{equation}
 m_t(P_{t,q,R})=R+1,\qquad
 |P_{t,q,R}|=q+\left\lceil\frac Rq\right\rceil+3,
 \label{eq:cerny-dictionary-count}
\end{equation}
where
\begin{equation}
 b_t=\begin{cases}5,&t=1,\\2t+2,&t\geq2.\end{cases}
 \label{eq:cerny-block-length}
\end{equation}
In particular, \(P_{t,q,2t}\) has first collision power \(2t+1\)
for every \(q\geq\kappa_t\).
\end{theorem}

\begin{proof}
Put \(\sigma=\lceil R/q\rceil\leq\lfloor q/2\rfloor\).
Lemma~\ref{lem:cerny-pair-spectrum} supplies the pair
\[
 U_R=\begin{cases}
 U_{\sigma q-R,\sigma},&q\text{ odd or }\sigma<q/2,\\
 H_{\binom q2-R},&q\text{ even and }\sigma=q/2,
 \end{cases}
\]
whose synchronization time is \(R\). Write \(U_R=\{u,v\}\) and
fix a state \(h\). Define \(\mathcal A_{q,R}\) on \(\mathcal Q_q\)
with labels \(\mathtt s,\mathtt a,\mathtt c\) and arcs
\[
 h\xrightarrow{\mathtt s}u,\quad h\xrightarrow{\mathtt s}v,\quad
 i\xrightarrow{\mathtt a}a(i)\ (i\in\mathcal Q_q),\quad
 i\xrightarrow{\mathtt c}c(i)\ (i\in J_\sigma),
\]
where \(J_\sigma=\{q-1,0,1,\ldots,\sigma-1\}\).
The successive \(c\)-steps in
\eqref{eq:cerny-nondiameter-descent} or
\eqref{eq:cerny-diameter-descent} act on
\(\{q-1,\sigma-1\},\ldots,\{q-1,0\}\). Thus the shortest
synchronizing word uses only retained arcs. Prefixing the two start arcs
gives two distinct walks of length \(R+1\) with the same endpoints and label word.

Conversely, choose a shortest pair of distinct walks with the same endpoints
and label word. These walks diverge at their first step and meet again only
at their last: a common prefix or an earlier meeting would give a shorter
such pair. The \(\mathtt a\)- and
\(\mathtt c\)-transitions are deterministic, so divergence uses the
two \(\mathtt s\)-arcs. A later \(\mathtt s\)-step would require
both walks to have returned to \(h\) before their final meeting.
The remaining word therefore uses only \(\mathtt a,\mathtt c\)
and synchronizes \(U_R\) in the full automaton, taking at least \(R\)
steps. Hence \(\operatorname{amb}(\mathcal A_{q,R})=R+1\).

Choose distinct guard symbols for the three labels, and take \(L_t=3\) when \(t=1\)
and \(L_t=2t\) when \(t\geq2\); for \(t=1\), \(L=2\) would make the second quantity in
\eqref{eq:guard-length-conditions} equal to \(2t=2\), rather than strictly larger.
The two quantities in
\eqref{eq:guard-length-conditions} are \(4,4\) for \(t=1\), and
\(2t+1,t(t+1)\) for \(t\geq2\), so both exceed \(2t\).
Theorem~\ref{thm:guarded-labelled-arc-transfer} gives
\(m_t(P(\mathcal A_{q,R},L_t))=R+1\). Set
\(P_{t,q,R}=P(\mathcal A_{q,R},L_t)\).
There are \(q\) cycle arcs, \(\sigma+1\) retained \(c\)-arcs, and
two start arcs. Their distinct encodings all have length \(L_t+2\),
proving the stated parameters.
\end{proof}

\begin{corollary}[Realizable first collision powers]
\label{cor:realizable-collision-powers}
For \(t\geq1\) and \(q\geq2\), every integer
\[
 1\leq a\leq\min\left\{2t+1,\binom q2+1\right\}
\]
is the first collision power of a block set over \(\Sigma_q\).
In particular, \(D_t(q)\geq\min\{2t+1,\binom q2+1\}\).
\end{corollary}

\begin{proof}
The sets \(\{01,10\}\) and \(\{0,1\}\) have first collision powers
one and two. This covers the asserted range for \(q=2\). For
\(q\geq3\) and \(a\geq3\), use
Theorem~\ref{thm:cerny-guarded-construction} with \(R=a-1\).
\end{proof}

\begin{example}[Explicit radius-two realization]
\label{ex:cerny-t2}
Take \(q=4\), \(L=4\), guards
\((g_{\mathtt s},g_{\mathtt a},g_{\mathtt c})=(0,1,2)\), and
\(h=0\). The pair \(\{0,1\}=U_{0,1}\) is synchronized by
\(\mathtt{aaac}\) in four steps. The eight blocks are
\[
 p^{\mathtt s}_0=0\,0^4\,0,\quad
 p^{\mathtt s}_1=0\,0^4\,1,\quad
 p^{\mathtt a}_i=i\,1^4\,a(i)\ (0\leq i<4),\quad
 p^{\mathtt c}_0=0\,2^4\,0,\quad p^{\mathtt c}_3=3\,2^4\,0.
\]
Interleaving the walks that start with
\(0\xrightarrow{\mathtt s}0\) and \(0\xrightarrow{\mathtt s}1\) and
then follow \(\mathtt{aaac}\), as in the proof of
Theorem~\ref{thm:guarded-labelled-arc-transfer}, gives the predecessors
\begin{align*}
 x&=p^{\mathtt s}_0p^{\mathtt a}_1p^{\mathtt a}_1p^{\mathtt a}_3p^{\mathtt c}_3
   =000000\,111112\,111112\,311110\,322220,\\
 z&=p^{\mathtt s}_1p^{\mathtt a}_0p^{\mathtt a}_2p^{\mathtt a}_2p^{\mathtt c}_0
   =000001\,011111\,211113\,211113\,022220.
\end{align*}
These predecessors differ only at their four reversed boundary pairs, so swapping
any two of these boundaries in \(x\) and the other two in \(z\) gives a common
received word. Applying \(\tau_6,\tau_{12}\) to \(x\) and \(\tau_{24},\tau_{18}\) to
\(z\) gives \(y=000001011111211112311110322220\). All four swaps exchange unequal symbols, and their
unordered pairs \(\{0,1\},\{1,2\},\{2,3\},\{0,3\}\) are distinct. The construction
theorem shows that the first four powers are \(2\)-correcting, so
this collision first occurs in power five.
\end{example}

\subsection{The alphabet obstruction}

The construction proves sufficiency in
Theorem~\ref{thm:exact-alphabet-threshold}. For the converse, in any collision
at first collision power \(2t+1\), a component that contains a differing block, in the
multigraph \(G\) of the proof of Theorem~\ref{thm:five}, must span
every block, since restriction to a smaller
component would contradict minimality. Connecting all blocks then uses all
\(2t\) available swaps, forcing one crossing per boundary. In the following
lemma, shortest predecessor paths to the received word ensure that the
swapped boundary pairs have unequal symbols; the proof then shows that
their unordered symbol pairs are distinct.

\begin{lemma}[First collisions at power \(2t+1\)]
\label{lem:maximal-witness}
Let \(t\geq1\) and \(P\subseteq\Sigma_q^b\) satisfy
\(m_t(P)=m=2t+1\). Then \(b\geq2\). In any collision
\(x\to y\leftarrow z\) in \(P^m\), with each predecessor path shortest
to the chosen \(y\), each path uses exactly \(t\) swaps, and each of the \(2t\) block boundaries is
crossed exactly once across the two paths. The words \(x,z\) agree away
from the boundary coordinates and have reversed unequal pairs \(u_i\mid v_i\) and
\(v_i\mid u_i\) at boundary \(i\). The unordered pairs
\(c_i=\{u_i,v_i\}\), \(1\leq i\leq2t\), are pairwise distinct.
\end{lemma}

\begin{proof}
If \(b=1\), a singleton \(P\) has no collision, while a set containing
distinct \(\alpha,\beta\) has first collision power two, witnessed by
\(\alpha\beta,\beta\alpha\). Thus \(b\geq2\).
Form the multigraph \(G\) of the proof of Theorem~\ref{thm:five} for the
chosen paths. A component containing a differing block must contain all
\(m\) blocks, since otherwise restriction gives a collision in a
smaller power. Connecting these blocks requires all \(m-1=2t\)
boundary edges, while the two paths together use at most \(2t\) swaps.
Hence each path has exactly \(t\) swaps, every swap crosses a block boundary,
and every boundary is crossed once.

Since \(b\geq2\), the coordinate pairs at the block boundaries are disjoint. Their swaps
commute, and the sets of boundaries crossed on the two paths are complementary. From
\(y=T_x(x)=T_z(z)\) we obtain \(z=T_zT_x(x)\), the result of
swapping the symbols at every block boundary of \(x\). Here shortestness
excludes equal-symbol swaps, which would be removable unchanged steps,
so the swapped pairs are unequal. This proves the
claimed form of \(x,z\).

Suppose that \(c_i=c_j\) for \(i<j\). If
\((u_i,v_i)=(u_j,v_j)\), delete blocks \(i+1,\ldots,j\) from
both words. If \((u_i,v_i)=(v_j,u_j)\), cross-splice instead:
\[
 \widetilde x=x^{(1)}\cdots x^{(i)}z^{(j+1)}\cdots z^{(m)},\qquad
 \widetilde z=z^{(1)}\cdots z^{(i)}x^{(j+1)}\cdots x^{(m)}.
\]
After deletion, the new boundary pairs are
\(u_i\mid v_j=u_i\mid v_i\) and \(v_i\mid u_j=v_i\mid u_i\);
after cross-splicing, they are
\(u_i\mid u_j=u_i\mid v_i\) and \(v_i\mid v_j=v_i\mid u_i\).
In both operations, retained boundary pairs remain reversed, with the word
roles exchanged on the cross-spliced suffix, and all other coordinates agree. Thus \(\widetilde x,\widetilde z\) are
distinct words of \(P^r\), where \(2\leq r=m-(j-i)<m\), differing
only at the coordinate pairs at their \(r-1\) block boundaries. Partition those boundaries into two
sets whose sizes differ by at most one, and swap the symbols at the boundaries in one set in
\(\widetilde x\) and those in the other set in \(\widetilde z\). The outputs
coincide, using at most \(\lceil(r-1)/2\rceil\leq t\) swaps on
each path. This contradicts the first collision power \(m\).
\end{proof}

\begin{proof}[Proof of Theorem~\ref{thm:exact-alphabet-threshold}]
If \(D_t(q)=2t+1\), attainment of the supremum in \eqref{eq:def-depth} gives a block set
with \(m_t(P)=2t+1\). Lemma~\ref{lem:maximal-witness} supplies
\(2t\) distinct unordered symbol pairs, so \(\binom q2\geq2t\).
Conversely, this inequality implies \(q\geq\kappa_t\geq3\), and
Theorem~\ref{thm:cerny-guarded-construction} with \(R=2t\) gives
\(D_t(q)\geq2t+1\). The reverse inequality is
Theorem~\ref{thm:five}. Solving \(\binom q2\geq2t\) for the least
integer \(q\) gives the stated value of \(Q_t\).
\end{proof}

\begin{corollary}[One pair below the threshold]
\label{cor:one-pair-deficit}
Let \(t\geq1\) and \(q\geq2\). If \(\binom q2=2t-1\), then
\(D_t(q)=2t\).
In particular, \(D_1(2)=2\), \(D_2(3)=4\), \(D_8(6)=16\), and
\(D_{11}(7)=22\).
\end{corollary}

\begin{proof}
Corollary~\ref{cor:realizable-collision-powers} gives the lower bound.
The finite-locality bound and Theorem~\ref{thm:exact-alphabet-threshold}
give \(D_t(q)<2t+1\), hence \(D_t(q)\leq2t\).
\end{proof}

At radius one, therefore, \(Q_1=3\):
\(D_1(2)=2\), whereas \(D_1(q)=3\) for every \(q\geq3\).
At radius two, the corresponding threshold is \(Q_2=4\), with
\(D_2(3)=4\) and \(D_2(q)=5\) for \(q\geq4\).
For \(\binom q2\leq2t-2\), these results give
\(\binom q2+1\leq D_t(q)\leq2t\), by Corollary~\ref{cor:realizable-collision-powers},
Theorem~\ref{thm:five} and Theorem~\ref{thm:exact-alphabet-threshold}; at
radius two this gives \(2\leq D_2(2)\leq4\).

\section{Explicit binary block codes}
\label{sec:moments}
\label{sec:correcting-block-sets}

The finite-locality theorem reduces correction at every concatenation length
to one finite test. We now give an explicit binary family, the derivative family, that corrects at
every concatenation length, with an independent block encoder and a decoder that recovers
block \(j\) from the first \(jb\) received symbols. Its construction uses the prefix-sum reduction and standard
binary BCH codes, and Proposition~\ref{prop:constituent-code-transfer} proves its correction directly
from their Hamming distance.

\subsection{Binary constituents and the prefix-sum map}

The established prefix-sum reduction turns each binary adjacent
transposition into at most one substitution
\cite{gabrys2018damerau,kovacevic2026bounds}. It therefore allows standard codes correcting
substitution errors to provide block sets for our channel. For
\(x,c\in\F_2^d\), define the linear prefix-sum (integral) map \(I_d\) and its inverse, the
derivative \(D_d\), by
\begin{equation}
 I_d(x)_i=\sum_{h=1}^{i}x_h,
 \qquad D_d(c)_i=c_i+c_{i-1},\qquad c_0=0.
 \label{eq:binary-integral-derivative}
\end{equation}
All sums in these maps are in \(\F_2\), and \(d_H\) denotes Hamming
distance. A swap at positions \(i,i+1\)
changes only coordinate \(i\) of \(I_d(x)\). Chaining this fact along any successive swap path from \(x\) to
\(z\) gives
\begin{equation}
 d_H\bigl(I_d(x),I_d(z)\bigr)\leq\AT(x,z)
 \qquad\text{for all }x,z\in\F_2^d.
 \label{eq:integral-distance-bound}
\end{equation}

Throughout, decoders are analyzed under the assumption that the received word
results from at most \(t\) successive adjacent transpositions of some codeword.
For the streaming cost analysis, the receiver is given the transmission
length \(n=mb\) before decoding starts.
Recovery positions count the received symbols needed; the computation
performed after those symbols arrive is accounted for separately.
Remark~\ref{rem:streaming} collects these receiver conventions in one place.

\begin{proposition}
\label{prop:constituent-code-transfer}
Let \(\mathcal B\subseteq\F_2^b\) be nonempty with Hamming distance at
least \(2t+1\), and set \(P=D_b(\mathcal B)\). Then \(|P|=|\mathcal B|\),
and every \(\cC_{mb}(P)\) corrects \(t\) successive adjacent
transpositions. An encoder and a radius-\(t\) Hamming decoder for
\(\mathcal B\) give an independent block encoder and a decoder that
recovers block \(j\) from the first \(jb\) received symbols, under the
stated error bound. If \(\mathcal B\) is closed under complement, then
\begin{equation}
 I_{mb}\bigl(\cC_{mb}(P)\bigr)=\mathcal B^m.
 \label{eq:complement-product-identity}
\end{equation}
\end{proposition}

\begin{proof}
The map \(D_b\) is bijective, giving the cardinality assertion.
Fix \(m\geq1\). For distinct \(x,z\in P^m\), let block \(j\) be the first where they differ. Their preceding blocks are identical, leaving a common cumulative parity \(a \in \F_2\) prior to block \(j\). Because \(I_b\circ D_b\) is the identity map, the global prefix sums \(I_{mb}(x)\) and \(I_{mb}(z)\) restricted to block \(j\) evaluate to \(a\mathbf1+c\) and \(a\mathbf1+c'\) for distinct \(c,c'\in\mathcal B\); these global prefix-sum segments thus have the same incoming parity offset \(a\). Since adding \(a\mathbf1\) preserves Hamming distance, they retain the constituent code's Hamming separation and differ in at least \(d_H(c,c') \geq 2t+1\) coordinates. Thus, \(\AT(x,z)\geq 2t+1\) by \eqref{eq:integral-distance-bound}, and Lemma~\ref{lem:distance} ensures \(\cC_{mb}(P)\) is \(t\)-correcting.

To encode, encode each message block as \(c\in\mathcal B\) and transmit
\(D_b(c)\). To decode \(y\), form \(v=I_{mb}(y)\) with one running
parity bit of the received word. Keep the decoded source parity \(a\)
at the preceding block boundary, initially zero. Subtract \(a\mathbf1\)
from the next block of \(v\), Hamming-decode the result to
\(c\in\mathcal B\), output \(D_b(c)\), whose parity is \(c_b\), and
replace \(a\) by \(a+c_b\). Under the stated error bound,
\(v\) has at most \(t\) Hamming errors in total relative to the true
global prefix sum. Inductively the stored \(a\) is correct, so every
constituent decoding has at most \(t\) errors and recovers the next source
block. For each block \(j\), a swap across the next block boundary changes only
integrated coordinate \(jb\), which the Hamming decoder corrects using
the decoded incoming source parity without observing symbol \(jb+1\).
Under the stated error bound, one block buffer and the constituent
decoder's workspace suffice.

If \(\mathcal B\) is complement-closed, every block \(c+a\mathbf1\)
of a global prefix sum belongs to \(\mathcal B\), proving one inclusion
in~\eqref{eq:complement-product-identity}. Conversely, for
\(u\in\mathcal B^m\), the local prefix sum of block \(j\) of
\(D_{mb}(u)\) is its block \(u^{(j)}\) plus
\(u_{(j-1)b}\mathbf1\), with the boundary bit taken as zero when \(j=1\). This too belongs to \(\mathcal B\), proving the
other inclusion. In this case one may Hamming-decode the blocks of \(v\)
independently and apply \(D_{mb}\) to their concatenation, using the
preceding decoded integrated bit at each block boundary.
\end{proof}

For exact acceptance on arbitrary inputs, require each constituent decoding
to return a word of \(\mathcal B\) or report failure, and accept a
reconstructed source only after verifying its adjacent-transposition
distance from the received word is at most \(t\).
On an input that is finally rejected, earlier emitted blocks carry no
recovery guarantee; withholding them until acceptance requires
buffering the reconstructed message.
For binary \(x,y\) of equal
weight, the representation of the distance by the positions of successive ones
\cite{kovacevic2019runlength} gives
\begin{equation}
 \AT(x,y)=\sum_{i=1}^{n-1}|d_i|,
 \qquad d_i=\wt(x_1\cdots x_i)-\wt(y_1\cdots y_i).
 \label{eq:binary-cut-distance}
\end{equation}
Since the prefix-composition difference of Lemma~\ref{lem:prefix-flow} has half
\(\ell_1\)-norm \(|d_i|\) for binary words, \eqref{eq:binary-cut-distance} is that
lemma holding with equality.
Indeed, match the \(k\)-th one of \(x\) to the \(k\)-th one of \(y\).
At cut \(i\), the crossing matched indices are precisely those above the
smaller prefix weight and at most the larger one; hence exactly \(|d_i|\)
matches cross, all in one direction. Summing counts their unit
displacements. Every swap of two unequal symbols moves a single one by one position,
giving the lower bound. For attainment, first move the ones whose matched
positions lie to their left in increasing order of their matched indices,
then those whose matched positions lie to their right in decreasing order.
In either phase, the order of the matched positions leaves only zeros
along each move, so the number of swaps is the total matched displacement. Unequal total weights give
infinite adjacent-transposition distance. A scan thus checks that \(d_n=0\) and the sum in
\eqref{eq:binary-cut-distance} is at most \(t\). Whenever a radius-\(t\)
predecessor exists, the decoder of
Proposition~\ref{prop:constituent-code-transfer} reconstructs it, and every accepted
candidate lies within \(t\) adjacent transpositions of the received
word.
For this final check, keep the current received block until its source block is reconstructed,
then update the global imbalance and accumulated distance by comparing
the two blocks before reusing their buffers.
Under the stated error bound, block \(j\) is recovered from the first
\(jb\) received symbols.

\subsection{An explicit family and its encoding and decoding costs}

Binary BCH codes provide constituents with logarithmic redundancy
\cite{bose1960binary}. We give
the defining checks and the needed parameter proof, so that the construction
and its decoder use the same coordinate convention.

\begin{lemma}[Binary BCH constituents]
\label{lem:binary-bch-parameters}
Let \(s,t\geq1\), \(L=2^s-1\), and \(2t<L\). Fix a primitive element
\(\alpha\in\F_{2^s}\), and define
\begin{equation}
 \mathcal B=\left\{c\in\F_2^L:
 \sum_{i=1}^{L}c_i\alpha^{h(i-1)}=0
 \quad(1\leq h\leq2t)\right\}.
 \label{eq:bch-checks}
\end{equation}
Then \(\mathcal B\) is a binary linear code of dimension at least
\(L-st\), Hamming distance at least \(2t+1\), and is closed under
complement. For every \(b\geq2t+1\), taking
\(s=\lceil\log_2(b+1)\rceil\) and shortening this code to length \(b\)
gives a binary linear code \(\mathcal B_b\) of dimension at least
\(b-st\) and the same distance lower bound. This bound guarantees
positive dimension when \(b>st\).
\end{lemma}

\begin{proof}
The check with index \(2h\) is the square of the check with index \(h\).
Repeated halving reduces every even index to one of the \(t\) odd indices
in \(\{1,3,\ldots,2t-1\}\). Each such field equation imposes at most
\(s\) binary linear constraints, so the dimension is at least \(L-st\).
For a nonzero word of weight \(w\leq2t\), let
\(\beta_1,\ldots,\beta_w\) be its distinct nonzero location values
\(\alpha^{i-1}\). Its first \(w\) checks say
\(\sum_{j=1}^w\beta_j^h=0\) for \(1\leq h\leq w\).
The \(w\times w\) matrix \((\beta_j^{h})_{1\leq h,j\leq w}\) is a
Vandermonde matrix times the invertible diagonal matrix with entries
\(\beta_j\), and hence is invertible.
It cannot
annihilate the all-one vector, a contradiction. Finally, \(\mathbf1\)
satisfies every check because \(1\leq h\leq2t<L\) and the geometric
sum of all \(L\) powers of \(\alpha^h\ne1\) vanishes. Linearity gives
complement closure.

For shortening, require coordinates \(b+1,\ldots,L\) to be zero and
delete them. These \(L-b\) constraints reduce dimension by at most
\(L-b\); deletion of prescribed-zero coordinates preserves dimension and
pairwise distance. The resulting dimension is at least \(b-st\).
\end{proof}

The shortened constituent need not be complement-closed, so we use the
sequential decoder in the proof of Proposition~\ref{prop:constituent-code-transfer},
carrying the decoded source parity between blocks. The following bounds are those of
the decoder specified and analyzed in the proof of Theorem~\ref{thm:explicit-binary-codec},
which follows the standard syndrome and linear-system method
\cite[Section~2.1, Proposition~2.1.2]{giacomelli2013decoding}.
For a nonnegative function \(g\), we write \(f=O(g)\) when there is a constant
\(C>0\) with \(|f|\leq Cg\) for all admissible parameter values, and
\(f=\Theta(g)\) when, in addition, there is a constant \(c>0\) with \(f\geq cg\) for
all these values; these constants do not depend on the parameters, except on those
named in a subscript or held fixed.
Stored code data consist of the polynomial defining \(\F_{2^s}\),
the primitive element, the systematic parity matrix, and the coordinate
ordering; powers of the primitive element are computed as needed.
Throughout, a bit operation is a Boolean operation on a bounded number of bits, the
reading or writing of one stored bit, or the movement of a head by one position on one
of a fixed number of sequential stores, so every scan and rewind is charged and the explicit families' bounds therefore hold without random access. Preparation
stores the parity matrix with one row for each parity coordinate, rows and entries in
increasing coordinate order, and the coordinate ordering as a mask of \(b\) bits marking
the information coordinates.

\begin{theorem}[Explicit binary block encoding and decoding]
\label{thm:explicit-binary-codec}
For \(t\geq1\) and \(b\geq2t+1\), set
\(s=\lceil\log_2(b+1)\rceil\). There is a binary linear block set
\(P\subseteq\F_2^b\) of dimension \(k\), with
\[
 r=b-k\leq st,
\]
such that every \(\cC_{mb}(P)\) corrects \(t\) successive adjacent
transpositions. After preparing a polynomial basis of \(\F_{2^s}\), a primitive element,
and a systematic constituent encoder, independent encoding of \(m\)
message blocks takes \(O(n(r+1))\) bit operations, where \(n=mb\).
Algebraic decoding takes
\begin{equation}
 O(nt+mt^3)\quad\text{operations in }\F_{2^s},
 \qquad O(b+t^2s)\quad\text{working bits},
 \label{eq:bch-decoder-cost}
\end{equation}
in addition to stream counters and stored code data. Under the stated error
bound, source block \(j\) is recovered upon receipt of symbol \(jb\).
With schoolbook arithmetic in a polynomial basis, the algebraic decoding cost is
\(O((nt+mt^3)s^2)\) bit operations.
\end{theorem}

\begin{proof}
Take \(P=D_b(\mathcal B_b)\), with \(\mathcal B_b\) from
Lemma~\ref{lem:binary-bch-parameters}. The dimension, redundancy, and
correction assertions follow from that lemma and
Proposition~\ref{prop:constituent-code-transfer}. In systematic form, the
constituent encoder computes its \(r\) parity coordinates from \(k\)
information coordinates using at most \(kr\) binary additions, each parity
coordinate in one pass over its stored row and the information bits; merging the
information and parity bits under the mask then restores the original coordinate
order in \(O(b)\) bit operations. Its parity matrix has \(kr\) bits, and the mask
has \(b\) bits. Differentiation costs \(O(b)\), giving the encoding bound.

It remains to specify and bound the constituent decoder. First reinsert the
\(L-b\) prescribed zeros, where \(L=2^s-1<2b+1\), obtaining
\(v\in\F_2^L\). Compute its syndromes and their Hankel matrix
\[
 S_h=\sum_{i=1}^{L}v_i\alpha^{h(i-1)}\quad(1\leq h\leq2t),
 \qquad H=(S_{a+c-1})_{a,c=1}^{t}.
\]
Find \(\nu=\operatorname{rank}H\) by Gaussian elimination. If \(\nu=0\),
set \(f(X)=1\). Otherwise solve
\begin{equation}
 \sum_{j=0}^{\nu-1}a_jS_{h+j}=S_{h+\nu}
 \quad(1\leq h\leq\nu),
 \qquad f(X)=X^\nu+\sum_{j=0}^{\nu-1}a_jX^j,
 \label{eq:bch-locator-system}
\end{equation}
rejecting if this system is singular. Evaluate \(f\) at every
\(\alpha^{i-1}\), require exactly \(\nu\) roots among these locations,
and flip the corresponding bits of \(v\). Return the shortened word only
if all defining syndromes vanish, the prescribed-zero coordinates remain
zero, and at most \(t\) bits were flipped; otherwise reject.

To verify recovery, suppose \(v=c+e\), where \(c\in\mathcal B\)
and \(e\) has weight \(w\leq t\). At its distinct error locations
\(\beta_1,\ldots,\beta_w\), the binary error values equal one, so
\(S_h=\sum_{j=1}^w\beta_j^h\). Consequently,
\[
 H=V\operatorname{diag}(\beta_1,\ldots,\beta_w)V^{\mathsf T},
 \qquad V_{a,j}=\beta_j^{a-1}\quad(1\leq a\leq t).
\]
The first \(w\) rows of \(V\) form an invertible Vandermonde matrix.
Thus \(V^{\mathsf T}\) is onto and
\(V\operatorname{diag}(\beta_j)\) is injective, giving
\(\nu=w\). For \(w>0\), the leading \(w\)-by-\(w\) submatrix
of \(H\) is also invertible. The monic polynomial
\(g(X)=\prod_{j=1}^w(X-\beta_j)\) satisfies
\eqref{eq:bch-locator-system}: multiply each equality \(g(\beta_j)=0\)
by \(\beta_j^h\) and sum. Uniqueness of the solution gives \(f=g\),
so precisely the erroneous bits are flipped. If \(w=0\), all syndromes
vanish and the decoder makes no flips. Conversely, every returned word
passes the code checks and is within Hamming distance \(t\) of its input.
This proves exact constituent decoding, including rejection outside its
Hamming balls. The preceding proposition supplies the blockwise channel
decoder; the check in~\eqref{eq:binary-cut-distance} supplies exact final
acceptance for adjacent transpositions.

Syndrome computation, root evaluation, and final syndrome checking each take
\(O(Lt)\) field operations, using successive powers or Horner evaluation.
One rank computation and one linear solve take \(O(t^3)\) field operations
and store \(O(t^2)\) field elements. A block buffer has \(O(b)\) bits,
and \(L=O(b)\), proving~\eqref{eq:bch-decoder-cost}. Integration,
differentiation, and comparison for the final distance check take \(O(n)\)
additional binary or counter operations.

In a supplied polynomial basis, multiplication and reduction cost
\(O(s^2)\) bit operations. Inversion by exponentiation to \(2^s-2\)
uses \(O(s)\) multiplications and costs \(O(s^3)\) bit operations. On the sequential
stores, \(v\), the rows of \(H\) in order, and the coefficients of \(f\) each have
their own store, with \(s\) consecutive bits per field element; a pivot step
copies the pivot row to a scratch store and rewinds this copy for each other row.
Head movement is then dominated by the accompanying field arithmetic: a syndrome scan moves \(O(L)\)
positions, a pivot step of the elimination \(O(t^2s)\), and a rewind of the locator
polynomial \(O(ts)\). Gaussian
elimination uses \(O(t)\) pivot inversions per block, so the bit cost is
\[
 O\bigl((nt+mt^3)s^2+mt s^3\bigr)
   =O\bigl((nt+mt^3)s^2\bigr),
\]
where the last equality uses \(s\leq L=O(b)\).
\end{proof}

Preparation is reusable for every \(m\). Once a polynomial basis of \(\F_{2^s}\) and
a primitive element are supplied, the remaining preparation expands the
checks over \(\F_2\) and applies binary Gaussian elimination to compute a
systematic generator in polynomial time in \(b,t,s\).
Let \(\Gamma(b,t)\) be the bit cost of the whole preparation, which supplies a polynomial
basis and a primitive element, computes the systematic encoder, and fixes the
layouts of the stored code data. Since the field data can be found by a finite
search, \(\Gamma(b,t)\) is finite and depends only on \(b\) and \(t\), so preparation is
independent of \(n\). We call the construction explicit because it is deterministic and
fully specified, with this finite preparation. Moreover, \(\Gamma(b,t)\) is polynomial in \(b\):
since \(2^s<2(b+1)\), the field data come from a deterministic search over fewer than
\(2(b+1)\) candidate defining polynomials and fewer than \(2b+1\) field elements, each tested
in time polynomial in \(b\), and the remaining preparation is polynomial in \(b,t,s\), where
\(t\leq(b-1)/2\). Including
preparation, the final distance check, and the framing procedure described below,
the first message costs \(\Gamma(b,t)+O(n(r+1))\) bit operations to encode and
\(\Gamma(b,t)+O\bigl((nt+mt^3)s^2+n\log(t+2)\bigr)\) to decode. These totals hold for
all \(t\geq1\) and \(b\geq2t+1\), with the displayed dependence on \(b\), \(t\), and
\(s=\lceil\log_2(b+1)\rceil\); for fixed \(b,t\), both are \(O(n)\) over the complete
transmission. The decoder uses \(O(b+t^2s)\) working bits, and the stored
code data, including the parity matrix of \(kr\) bits and the mask of \(b\) bits,
are reused. Framing adds \(O(\log n)\) bits, and output storage is additional when
release waits for final acceptance.
For final distance checking, the accumulated distance can be capped
at \(t+1\) and rejection declared immediately when it exceeds \(t\);
before rejection the imbalance counters are bounded by \(t+1\).
Thus this check adds \(O(n\log(t+2))\) bit operations. For framing, use a binary countdown from \(n\) and a block-position
counter. The countdown uses \(O(n)\) bit operations over the whole
transmission: decrementing it touches
\(n+\sum_{j\geq1}\lfloor n/2^j\rfloor<2n\) bits in total,
with constant-factor overhead for returning to its low-order bit.
The block-position counter has size bounded in terms of \(b\).
The test for the end of the frame compares the countdown with zero through a mark on
its highest nonzero bit, which the decrements maintain, as Remark~\ref{rem:streaming}
describes; the test for a block boundary reads the block-position counter, at
\(O(\log b)\) bit operations per symbol. Framing therefore takes \(O(n\log b)\) bit
operations, and \(O(n)\) for fixed \(b,t\). Both totals above absorb this: for
decoding because \(\log_2b<s\), and for encoding because the Hamming packing bound,
applied to the constituent \(\mathcal B_b\), which has length \(b\), distance at least
\(2t+1\geq3\) and the same redundancy \(r\), gives \(r\geq\lceil\log_2(b+1)\rceil\).

The code carries \(mk\) information bits and has total redundancy
\(mr\leq mt\lceil\log_2(b+1)\rceil\). At fixed \(t\), its rate
tends to one as \(b\) grows. At fixed \(b,t\), increasing \(m\)
reuses the same encoder and decoder while leaving the rate fixed. 
Under independent block encoding, the total redundancy \(mr\) grows linearly with \(m\),
whereas the same prepared code data and bounded workspace are reused as \(m\)
grows. A single prepared implementation therefore encodes and decodes finite framed
transmissions of any length \(n=mb\) known to the receiver.
Every binary block set correcting at every concatenation length must itself
correct \(t\) errors at length \(b\). Kova\v{c}evi\'c's upper bound \(O_t(2^b/b^t)\) on the
size of ordinary binary codes
\cite[Theorem~11, Corollary~12, and equation~(54)]{kovacevic2019runlength} covers
such codes. It concerns codes correcting a total displacement of the ordered ones of
at most \(t\), whose codewords may have any weight and, by the convention stated
right after his equation~(54), may end in either bit. For equal weights this
displacement equals the distance in~\eqref{eq:binary-cut-distance}, and words of
unequal weights are infinitely far apart under both distances. The bound therefore
gives per-block redundancy at least \(t\log_2 b-O_t(1)\),
so this family's bound \(t\lceil\log_2(b+1)\rceil\) is optimal up to
an additive constant for fixed \(t\).

\section{Decoding delay}
\label{sec:blockwise-decoding}

The explicit binary family in the preceding section permits the same encoder to be
used independently on every message block. We now analyze the decoding delay required
to recover a block. Received symbols arrive sequentially from left
to right, and the decoder knows the block parameters \(P,b,t\) and the total sequence length \(n=mb\).
The composition of the transmitted word remains unknown to the decoder.
The additional decoding delay for block \(j\) is defined as the number of received symbols
needed beyond position \(jb\) to determine that block. That is, the symbols beyond position \(jb\) are those at
positions \(jb+1,jb+2,\ldots\), position \(jb\) being the last position
of block \(j\). This measures
delay by symbol arrivals; computational operations performed between arrivals are
accounted for separately.

Correct and irrevocable block recovery is required for every word within \(t\)
successive adjacent transpositions of \(P^m\), with a single swap budget for
the entire word. On inputs outside this radius, the decoders below can report failure
after emitting a prefix. An interface that releases output only after 
final channel acceptance buffers emitted blocks until the stream terminates. Working-memory bounds below exclude this final output buffer,
as well as static, reusable code descriptions.

\subsection{The general delay bound}

Two possible transmissions that remain consistent with a received prefix
agree on every source block that ends at least \(2tb\) symbols before the end
of the prefix. Their two swap paths
together cross at most \(2t\) block boundaries. An untouched boundary
therefore lets us apply correction to a shorter pair of source prefixes.

\begin{theorem}
\label{thm:block-decoding-delay}
Let \(P\subseteq\Sigma_q^b\) be nonempty, and suppose that
\(\cC_{mb}(P)\) corrects \(t\) successive adjacent transpositions.
For every \(1\leq j\leq m\), all transmissions within the error bound
whose received words share their first \(\min\{n,(j+2t)b\}\) symbols
have the same first \(j\) source blocks, even if their received words
differ later.
\end{theorem}

\begin{proof}
First, \(P^h\) is \(t\)-correcting for every \(h\leq m\).
A collision in \(P^h\) would extend to one in \(P^m\) by appending
the same \(m-h\) blocks of \(P\) to both sources and their common received
word. At position \(n\), the claimed determination follows directly
from correction of \(P^m\).

Suppose instead that \((j+2t)b<n\). Consider two sources in \(P^m\)
and two paths of at most \(t\) swaps whose received words agree through
position \((j+2t)b\). Among the \(2t+1\) cuts
\[
 jb,(j+1)b,\ldots,(j+2t)b,
\]
at least one is crossed by neither path. Indeed, each crossing uses a
swap at that cut, and the two paths contain at most \(2t\) swaps in
total. Restrict both paths to
the prefix ending at an untouched cut \(hb\). The source prefixes
belong to \(P^h\), and their received prefixes agree. Correction of
\(P^h\) forces the source prefixes to agree. Since \(h\geq j\),
their first \(j\) blocks agree.
\end{proof}

Thus at most \(2tb\) additional received symbols beyond position $jb$ suffice to decode source block $j$, for every
alphabet and every correcting block set. The finite-locality result of
Section~\ref{sec:model} concerns a different quantity: \(2t+1\) is
the number of source blocks needed to certify correction at every
concatenation length. Here correction is assumed, and the bound describes
the decoding delay during stream processing when source blocks become uniquely determined.

\subsection{Binary decoding with bounded working memory}

The preceding theorem gives unique determination over any alphabet.
We now give a general decoder, using~\eqref{eq:binary-cut-distance}, for binary block
sets whose layered recognizer, defined below, is supplied with the set or has width
small compared with \(|P|\). It serves block sets without a decoder of their own;
the explicit families of Theorem~\ref{thm:explicit-binary-codec} and
Corollary~\ref{cor:bch-high-rate-delay} have their own decoders. We assume a
deterministic layered recognizer for \(P\): its layers correspond to positions \(0,\ldots,b\) in one
block, its edges read one binary symbol, and acceptance at layer \(b\)
resets it to the initial state for the next block. Let \(W\) bound the
number of states in any layer. Trimming states outside accepting paths and merging the
accepting states at layer \(b\) yield a proper trellis of \(P\) without increasing the width.
A trie gives such a recognizer, of width at most \(|P|\),
for a listed block set; a more compact description can give a smaller width.

Proposition~\ref{prop:binary-streaming-decoder} counts operations in the word RAM
model rather than the bit operations of Section~\ref{sec:correcting-block-sets}:
addition, subtraction, multiplication, shifts, comparison, conditional branching,
and direct or indirect memory access on words each count as one operation, as do
reading and emitting one bit. Words accommodate recognizer-state indices, stream
positions, addresses of the working memory, and integers of magnitude \(O(t)\),
such as costs, imbalances, their indices, and the product in the completion test of
the proof below. Their bit lengths therefore depend on
\(W,t,b,n\); representing an arbitrary position from \(0\) through \(n\)
requires at least \(\lceil\log_2(n+1)\rceil\) bits. The width \(W\) bounds
one layer, while construction and storage of the full recognizer are
separate resources. The proposition assumes that a transition for a
specified layer, state, and input bit is accessible in \(O(1)\) word
operations, with the stored recognizer addressable in this model.

\begin{proposition}
\label{prop:binary-streaming-decoder}
Let \(P\subseteq\{0,1\}^b\) be nonempty and have a deterministic
layered recognizer of width at most \(W\), with constant-time
transitions. If \(\cC_{mb}(P)\) corrects \(t\) successive adjacent
transpositions, it has a decoder using
\[
 O\bigl(nW(t+1)\bigr)\quad\text{word operations},\qquad
 O\bigl(bW(t+1)^2\bigr)\quad\text{working words}.
\]
Block \(j\) is output upon reaching received position
\(\min\{n,(j+2t)b\}\). Recognizer construction and storage, and
storage of the input and emitted output, are separate resources.
\end{proposition}

\begin{proof}
For a candidate source prefix \(a_1\cdots a_i\) and the observed
prefix \(y_1\cdots y_i\), put
\[
 d_h=\wt(a_1\cdots a_h)-\wt(y_1\cdots y_h),\qquad
 c=\sum_{h=1}^{i}|d_h|.
\]
The integer \(d_i\) records the excess source weight in the prefix;
\(c\) is the cost already accumulated. The binary distance identity
\eqref{eq:binary-cut-distance}, equivalently the shift-distance representation
by the positions of successive ones~\cite{kovacevic2019runlength}, gives
\(\AT(a,y)=c\) at the end precisely when \(d_n=0\).

After reading position \(i\), retain for each recognizer state \(s\)
and imbalance \(d\) only the smallest attainable cost \(c\),
together with one predecessor pointer and its source bit. Initially the
recognizer is at its initial state and \(d=c=0\). A source
bit \(a\in\{0,1\}\) with a transition from \(s\) gives the update
\begin{equation}
 d'=d+a-y_{i+1},\qquad c'=c+|d'|.
 \label{eq:binary-prefix-recurrence}
\end{equation}
Block acceptance and reset are checked at every multiple of \(b\).
For two histories with the same \(s,d\), each possible source
continuation has the same recognizer path from \(s\) and the same future
cost increments. A minimum-cost history can therefore replace the other
without losing an accepting continuation. This is a weighted
shortest-path calculation on a product automaton, as in the general
framework of~\cite[Section~5]{mohri2003edit}.

Early output requires an additional test. At position \(i\), set
\(e=|d|\). With the source and received suffixes ranging freely over
\(\{0,1\}^{n-i}\), completion at total distance at most \(t\) is
possible if and only if
\begin{equation}
 n-i\geq e,\qquad c+\frac{e(e-1)}2\leq t.
 \label{eq:binary-suffix-completion}
\end{equation}
To prove necessity, each new pair of bits changes the imbalance by at
most one. Reaching zero therefore takes at least \(e\) positions
and adds cost at least \((e-1)+\cdots+1\); the current contribution
\(|d|\) is already included in \(c\). For sufficiency, use
source/received pairs \((0,1)\) when \(d>0\), or \((1,0)\)
when \(d<0\), until the imbalance is zero. Fill any remaining
positions with equal bits. The distance identity then gives the claimed
completion. Discard every state failing~\eqref{eq:binary-suffix-completion}.
This test is monotone in \(c\), so minimum-cost replacement remains
valid. Every codeword predecessor in \(B_t^{\mathrm{AT}}(y)\) survives these steps
through some representative. At the end the test forces \(d=0\),
and the stored cost is its exact adjacent-transposition distance.

We next justify output before the end. At a block boundary \(kb\),
every retained history contains \(k\) blocks of \(P\) and, by the
completion test, extends to some full pair of binary words joined by at
most \(t\) swaps. The received extension agrees with the observed
prefix. For any two histories and \(k>2t\), apply the untouched-cut
argument to these extensions and the cuts
\[
 (k-2t)b,(k-2t+1)b,\ldots,kb.
\]
At the untouched cut both source prefixes are words of \(P^*\),
even though their arbitrary suffix extensions need not
be. Correction of the corresponding shorter power forces agreement
through block \(k-2t\). All retained histories consequently have
the same first \(k-2t\) blocks. Output the next common block and
discard its stored history. At the end, output all remaining blocks of
the unique accepted history, or report failure if none remains.
For an input within the stated error bound an accepting history always remains, so each
earlier output is a prefix of the unique source.

For the resource bound, \(c\leq t\) implies \(|d|\leq t\).
There are at most \(W(2t+1)\) cells at each position, with two
possible transitions per cell. Store predecessor nodes only for the
uncommitted part of the stream. Immediately before a block is output,
this part spans at most \((2t+1)b\) positions. Following any one
live predecessor chain reconstructs the next block in
\(O((t+1)b)\) operations. Remove the committed layers and cut the
predecessor pointers at the retained boundary; keep the accumulated costs,
so the swap budget continues to apply to the entire word. Retiring
nodes takes at most as many operations as creating them. These counts give
\(O(nW(t+1))\) operations and \(O(bW(t+1)^2)\) working words.
When the stream ends before a scheduled block output, its entire length
is already within this window. For \(t=0\), each block is checked
and output upon receipt of its last bit.
\end{proof}

The decoder reads the input once and
need not retain its past symbols. A listed block set can have a large
recognizer, whose preparation and stored representation remain outside
the online bounds.

\subsection{High-rate block sets attaining the bound}

The derivative construction in
Proposition~\ref{prop:constituent-code-transfer} has zero additional
decoding delay. We now construct block sets of rate tending to
one for which the full \(2tb\) delay is necessary. We use the binary
maps \(I_d,D_d\) from~\eqref{eq:binary-integral-derivative}.
A code with prescribed Hamming distance will determine the block
interiors, while the two endpoints permit uncertainty to persist
across successive blocks.

\begin{theorem}
\label{thm:high-rate-delay}
Let \(t\geq1\). Suppose that \(\mathcal B\subseteq\F_2^L\)
has Hamming distance at least \(2t+1\), is closed under complement,
and has at least four words. Put \(b=L+2\) and
\(\mathcal B_0=\{c\in\mathcal B:c_1=0\}\), selecting one word per
complement pair. Partition \(D_L(\mathcal B_0)=U_0\sqcup U_1\)
into any two nonempty sets, and define
\begin{equation}
 P=\{0ue:u\in U_0,\ e\in\F_2\}
   \cup\{eve:v\in U_1,\ e\in\F_2\}.
 \label{eq:high-rate-delay-blocks}
\end{equation}
Then \(|P|=|\mathcal B|\), and \(\cC_{mb}(P)\) corrects
\(t\) successive adjacent transpositions for every \(m\geq1\).
For every \(m\geq2t+1\), the maximum additional delay required for
unique determination, over source-block indices and valid received
words, is exactly \(2tb\).
\end{theorem}

\begin{proof}
Complementation pairs the words with first coordinate zero and one,
so \(|\mathcal B_0|=|\mathcal B|/2\), while \(|\mathcal B|\geq4\) makes the
partition into two nonempty classes possible. The derivative is
injective, the two interior classes are disjoint, and each interior
has two distinct endpoint choices. Hence \(|P|=|\mathcal B|\).

For correction, suppose that \(x,z\in P^m\) are joined by at most
\(2t\) swaps. By~\eqref{eq:integral-distance-bound}, their global
binary prefix sums have Hamming distance at most \(2t\). In a
corresponding block with interior \(D_L(c)\), the \(L\) coordinates
of the global prefix sum corresponding to that interior are
\(c+a\mathbf1\), where \(c\in\mathcal B_0\),
\(\mathbf1\) is the all-one word of length \(L\), and \(a\)
is the prefix parity immediately after the left endpoint. Complement
closure puts this segment in \(\mathcal B\). The corresponding
two segments for \(x,z\) must agree by the distance of
\(\mathcal B\). Since \(c_1=0\), the first coordinate of each segment \(c+a\mathbf1\)
equals its offset \(a\). Their first coordinates give equality of the
offsets \(a\), and then equality of \(c\). Thus every aligned
pair of interiors agrees, including its type \(U_0\) or \(U_1\).

It remains to determine the endpoint bits. The parity argument identified
the interiors and their parity offsets; we now use integer prefix weights
to obtain the endpoint relations. The total absolute imbalance bound
will force zero imbalance through each long interior and allow only
compensating changes at a block boundary. Put
\[
 h_i=\wt(x_1\cdots x_i)-\wt(z_1\cdots z_i).
\]
Each swap changes one integer prefix weight by one, so
\(\sum_{i=1}^{mb-1}|h_i|\leq2t\), and \(h_0=h_{mb}=0\).
Equal interiors make \(h_i\) constant on the \(L+1\) cuts from
just after a block's left endpoint to just before its right endpoint.

Complement closure gives \(L=d_H(c,\bar c)\geq2t+1\) for
every \(c\in\mathcal B\). Every such constant is consequently zero.
It follows that the first left endpoints agree, the last right
endpoints agree, and the endpoints at each block boundary satisfy
\begin{equation}
 (x_{kb}-z_{kb})+(x_{kb+1}-z_{kb+1})=0,
 \qquad 1\leq k<m,
 \label{eq:delay-seam-balance}
\end{equation}
where the differences are integers.

With equal interiors, a block of type \(U_0\), whose left endpoint
is fixed, permits a first endpoint disagreement, while a block of
type \(U_1\), whose endpoints are equal, propagates it.
If the words differ, their first differing block has equal left
endpoints, by the preceding boundary equation or the initial endpoint
equality, and equal interiors. Its right endpoints must therefore
differ, so its type is \(U_0\): type \(U_1\) would force them equal.
Equation~\eqref{eq:delay-seam-balance} carries any right-endpoint
disagreement to the next left endpoints. Each subsequent block must
therefore have type \(U_1\), since type \(U_0\) has fixed left
endpoint zero, and its equal-endpoint rule carries the disagreement
to its right endpoints. The disagreement thus reaches the final
right endpoints, contradicting their equality. Hence \(x=z\), and
Lemma~\ref{lem:distance} proves correction for every \(m\).

For the delay lower bound, choose \(u\in U_0\), \(v\in U_1\)
and abbreviate \(a_e=0ue\), \(w_e=eve\). Take the two sources
of \(2t+1\) blocks
\begin{equation}
 x=a_0(w_1w_0)^t,\qquad z=a_1(w_0w_1)^t.
 \label{eq:delay-lower-sources}
\end{equation}
At each of their \(2t\) block boundaries, the endpoint pairs are unequal-bit
reversals. Choose any set of \(t\) boundaries and swap the endpoint pairs there in \(x\);
swap the endpoint pairs at the complementary \(t\) boundaries in \(z\). These coordinate pairs
are disjoint, so these are two valid paths of exactly \(t\) swaps.
At a chosen boundary, the first pair becomes the second; at an unchosen
boundary, the second becomes the first. All interiors and the first left
endpoint already agree. Hence the resulting received words agree
through position \((2t+1)b-1\) and differ only at their last
coordinate, which is zero in the first word and one in the second.
Their first source blocks differ. A decoder with the stated side
information therefore cannot complete the first block before receiving
symbol \((2t+1)b\), giving \(2tb\) additional symbols.
For larger \(m\), append the same blocks of \(P\) to both sources and
received words. The two paths and their common received prefix remain
valid. Theorem~\ref{thm:block-decoding-delay} gives the matching upper
bound.
\end{proof}

The two sources in~\eqref{eq:delay-lower-sources} have weights differing
by one. Each channel path preserves its own source weight; since that weight
is not supplied to the decoder, both sources stay consistent with what it knows.
Appending at least one common block places the necessary
decision strictly before the end of the stream.
For frames of at least \(2t+2\) blocks, the same worst-case delay \(2tb\) occurs even when the
receiver is told the composition: append \(0u'1\) to the lighter source \(x\)
in~\eqref{eq:delay-lower-sources} and \(0u'0\) to \(z\), for any \(u'\in U_0\),
and then common blocks if needed, so the two source weights become equal while their
received words still agree before symbol \((2t+1)b\).

\begin{corollary}[The endpoint family]
\label{cor:bch-high-rate-delay}
Let \(L=2^s-1\), \(2t<L\), and \(L-st\geq2\).
The unshortened complement-closed BCH constituent of
Lemma~\ref{lem:binary-bch-parameters} gives an explicit binary block set \(P\) of
length \(b=L+2\) with redundancy at most
\[
 b-\log_2|P|\leq st+2=t\log_2(b-1)+2
\]
that has the properties stated in Theorem~\ref{thm:high-rate-delay}. For fixed \(t\),
its rate tends to one along these block lengths.
\end{corollary}

\begin{proof}
The primitive binary BCH code of Lemma~\ref{lem:binary-bch-parameters}
has dimension \(k\geq L-st\), distance at least \(2t+1\), and
contains the all-one word. It is therefore complement-closed and has
at least four words. Apply Theorem~\ref{thm:high-rate-delay}; its
cardinality formula gives \(|P|=2^k\) and \(b-k\leq st+2\).
Dividing this redundancy bound by \(b=2^s+1\) gives the rate limit.
\end{proof}

The specialization in Corollary~\ref{cor:bch-high-rate-delay} also has a direct block encoder and a decoder that
avoids listing \(P\) and releases block \(j\) by received position \(\min\{n,(j+2t)b\}\);
these explicit algorithms, whose costs follow, thus attain the smallest worst-case additional delay
that Theorem~\ref{thm:high-rate-delay} shows possible for this family. The subspace \(\mathcal B_0\) has dimension
\(k-1\). Choose a systematic encoder for it and let \(\ell(c)\)
be one of its information coordinates. Partition its derivatives
according to \(\ell(c)=0\) or \(1\). The resulting block set \(P\) is nonlinear,
with encoding map
\begin{equation}
 (c,e)\longmapsto\bigl(e\ell(c),D_L(c),e\bigr),
 \qquad c\in\mathcal B_0,\quad e\in\F_2.
 \label{eq:high-rate-delay-encoder}
\end{equation}
It uses \(k\) information bits per block. If \(r=L-k\),
the subspace \(\mathcal B_0\) has redundancy \(r+1\); a prepared
systematic parity matrix evaluates the map in \(O(b(r+2))\) bit
operations per block, including differentiation and endpoint output.

To decode, form \(I_n(y)\) with one running parity bit. In each
received block, Hamming-decode the \(L\) interior coordinates of
this global prefix sum to a word \(\widehat c\in\mathcal B\).
Under the stated error bound, this segment differs in at most \(t\)
coordinates from the true \(c+a\mathbf1\), where \(a\) is the prefix parity of the
source immediately after the left endpoint. Since \(c_1=0\),
\(\widehat c_1=a\). Normalizing by
\(c=\widehat c+\widehat c_1\mathbf1\) and applying \(D_L\)
therefore recovers the interior and its type before the preceding endpoints
have been fixed. The sequential decoder in the proof of
Proposition~\ref{prop:constituent-code-transfer} instead subtracts the decoded
prefix parity of the source at the preceding block boundary. Here \(\widehat c_1\)
combines that parity with the current left endpoint, so the two values of \(e\)
in~\eqref{eq:high-rate-delay-encoder} supply the remaining block candidates.

For each imbalance \(d\), retain a minimum prefix cost and a
predecessor pointer at block boundaries. Compare each of the two candidate
blocks with the received block, bit by bit, using
\eqref{eq:binary-prefix-recurrence}; discard a branch as soon as its
cost exceeds \(t\), and apply~\eqref{eq:binary-suffix-completion}
at the boundary. Equal imbalances have identical future candidate
blocks and future cost increments, since future interiors are recovered
from received prefix sums. Thus minimum-cost replacement is valid.
At a block boundary with imbalance \(d\), the prefix parity of the source is that of
the received word plus \(d\) modulo~\(2\), and it fixes each candidate's segment of the
global prefix sum. Since \(\widehat c\) is within Hamming distance \(t\) of the received
segment, a candidate with segment \(\widehat c+\mathbf1\) disagrees with it in at least
\(L-t\) integrated bits. A disagreeing integrated bit at position \(h\) makes the prefix
weights through \(h\) differ in parity, so \(d_h\) is odd and \(|d_h|\geq1\). Summed over
the interior, the candidate's prefix cost is at least \(L-t>t\), as \(2t<L\); this bound
depends only on the prefix imbalances. The cost test thus discards
every candidate with the wrong offset.
The true source survives, and every retained history consists of
blocks of \(P\) with an arbitrary radius-\(t\) completion. The same
untouched-cut argument permits commitment after \(2t\) further
blocks. At the end, zero imbalance and the accumulated cost certify
exact acceptance; a failed constituent decoding or an empty state set
causes rejection.

There are at most \(2t+1\) boundary imbalances and two length-\(b\)
candidate blocks per state. The endpoint calculation takes \(O(nt)\)
word operations. Store the common recovered interiors for the last
\(2t+1\) blocks, and store only endpoint choices and predecessor
pointers in the boundary nodes. This uses \(O(tb)\) buffered bits,
including the current received block, and \(O(t^2)\) working words for the boundary
nodes. Here prefix costs and imbalances are bounded by \(t+1\), and predecessor
pointers index these \(O(t^2)\) nodes, so costs, imbalances and pointers have
\(O(\log(t+2))\) bits; in the bit operations of Section~\ref{sec:correcting-block-sets},
the endpoint calculation therefore takes \(O_t(n)\) bit operations, separate from
constituent arithmetic and global framing.
A constituent Hamming decoder with time \(T(L,t)\) and workspace
\(S(L,t)\) adds \(mT(L,t)\) time and \(S(L,t)\) workspace in
its own cost model. The explicit constituent decoder in the proof of
Theorem~\ref{thm:explicit-binary-codec}, applied to the \(L\) interior
coordinates of the unshortened constituent, gives \(O(nt+mt^3)\) field operations in
\(\F_{2^s}\) as in~\eqref{eq:bch-decoder-cost}, in addition to the endpoint calculation. The
bit-cost conversion stated in that theorem and the reusable preparation
costs described after its proof apply unchanged.

With the BCH constituents specified above, the derivative family and the endpoint
family both have block redundancy
\(t\log_2 b+O_t(1)\) for fixed \(t\), while their worst-case
additional delays are zero and \(2tb\), respectively. For the lower
redundancy estimate, a constituent of length \(d\) and dimension \(k\)
satisfies the Hamming packing bound
\(d-k\geq\log_2\sum_{i=0}^t\binom di=t\log_2 d-O_t(1)\).
Apply this with \(d=b\) for the derivative family and \(d=b-2\) for
the endpoint family; the BCH bounds supply the other direction. Thus, as the block length grows, rate tending to one is compatible with
either zero additional delay or worst-case additional delay \(2tb\).
The endpoint family shows, through Theorem~\ref{thm:high-rate-delay}, that the delay bound of
Theorem~\ref{thm:block-decoding-delay} is sharp at rates tending to one. At these lengths,
Proposition~\ref{prop:constituent-code-transfer}, applied to the endpoint family's BCH
constituent with two zero coordinates appended, gives a block set of the same length and size
with zero additional delay.
The derivative family of Theorem~\ref{thm:explicit-binary-codec} instead, at \(b=2^s+1\),
shortens a BCH code of length \(2^{s+1}-1\) and guarantees redundancy at most
\(t\lceil\log_2(b+1)\rceil=t(s+1)>st+2\) when \(t\geq3\); the endpoint family's
smaller stated guarantee at these lengths therefore reflects the choice of constituent.

\begin{remark}[Streaming operation]
\label{rem:streaming}

These results describe a streaming receiver for a framed transmission
whose length \(n=mb\) is known. Preparation and stored code data are
reused for every \(m\), and block \(j\) is output by received position
\(\min\{n,(j+2t)b\}\), or upon receipt of symbol \(jb\) in the derivative family
(Theorem~\ref{thm:explicit-binary-codec}). The general recognizer-based decoder uses a
number of working words bounded independently of \(m\); their bit
lengths follow the word RAM model stated above.

For the two explicit BCH-based families, the stated error bound, exact acceptance on
arbitrary inputs, and early or withheld release of output are described before and after
Proposition~\ref{prop:constituent-code-transfer} and at the opening of
Section~\ref{sec:blockwise-decoding}, where recovery positions count symbol
arrivals; computational resources are stated after the proof of
Theorem~\ref{thm:explicit-binary-codec} and after
Corollary~\ref{cor:bch-high-rate-delay}, with preparation charged once. In both,
constituent decoding has the costs of Theorem~\ref{thm:explicit-binary-codec} in
field and bit operations. The derivative family adds integration,
differentiation, and the final distance check; the endpoint family adds the
\(O(nt)\) word operations of the endpoint calculation, which also cover its
integration, normalization, and differentiation and use operands and pointers of
\(O(\log(t+2))\) bits. For fixed \(b,t\), the local data-processing cost per
block is independent of \(m\), so the numbers of operations in \(\F_{2^s}\), bit
operations, and word operations are each linear in \(n\); for growing parameters,
each has the dependence on the parameters stated where it is derived.

For both explicit BCH-based families, the data memory is bounded in terms of \(b,t\),
with framing adding \(O(\log n)\) counter bits. The endpoint storage
described above has this same bit-memory interpretation. These bounds exclude
already emitted output and the reusable code description. The total
length is used in the output position \(\min\{n,(j+2t)b\}\) and in the completion test
\eqref{eq:binary-suffix-completion}; the delay bound \(2tb\) is
independent of \(n\). The countdown described in Section~\ref{sec:correcting-block-sets}
adds \(O(n)\) bit operations over the complete transmission; the endpoint
completion test compares the remaining length only with \(|d|\leq t\).
For comparisons with values at most \(t\), a mark on the highest nonzero bit
identifies whether the remaining length fits in
\(\lceil\log_2(t+1)\rceil\) bits; only in that case are those bits inspected.
The decrements already visit the bits needed to move the mark, and a flag recording
whether the length fits, set with the counter, changes only when the mark crosses
the threshold; each comparison thus uses \(O(\log(t+2))\) bit operations beyond
the countdown's amortized \(O(n)\).

\end{remark}

\section{Extraction from ordinary correcting codes}
\label{sec:extraction}

A code correcting adjacent transpositions at one fixed length can lose
correction when its words are concatenated. We seek one fixed subset,
with the same block length and ambient alphabet, that corrects at every
concatenation length, as the following example illustrates.

\begin{example}[Correction under concatenation]
\label{ex:composition-extraction}
Let \(t=1\), \(b=2\), and \(\cD=\{00,01,11\}\). Its three words
have different weights, so \(\cD\) corrects one adjacent transposition.
Its second power contains
\[
 x=00\,|\,11=0011,\qquad z=01\,|\,01=0101.
\]
Swapping positions \(2,3\) in \(x\) gives \(z\). Thus the received
word \(y=z\) has predecessors \(x\) and \(z\), reached with one and
zero swaps, respectively, and \(\cD^2\) fails to correct one error.
The subset \(P=\{00,11\}\), however, corrects at every concatenation
length; its blocks have different compositions but equal composition
residues modulo \(2\), illustrating the later modular criterion. Every swap of two unequal symbols crosses a boundary and changes either
\(00\,|\,11\) into \(01\,|\,01\), or \(11\,|\,00\) into
\(10\,|\,10\). The resulting two mixed blocks identify the boundary
and the original pair uniquely. A word received unchanged consists entirely
of blocks \(00\) and \(11\). Thus every word produced by at most one
swap has a unique predecessor in \(P^m\), and the restriction retains two of the three
available blocks.
\end{example}

For a fixed alphabet and error radius, we will show that every ordinary
correcting code contains such a subset retaining a constant fraction of its
words. We first treat equal block compositions, then pass to composition
residues with a modulus depending only on the error radius, giving a
bounded number of classes independent of the block length.

\subsection{Constant-composition blocks}

For a pair of words whose corresponding blocks have equal compositions,
the number of inversions of the matching that pairs the \(k\)-th
occurrences of each symbol separates across blocks, even
though swap paths may cross their boundaries. Equal cumulative symbol counts before
each block keep this matching inside corresponding block intervals, and its inversion
count is the adjacent-transposition distance, which is therefore additive over the
blocks. A common composition throughout the block set supplies this pairwise
condition for every pair in every power.

\begin{proposition}[Constant-composition blocks]
\label{prop:constant-composition}
Suppose that \(P\subseteq\Sigma^b\) is nonempty and all its blocks have
the same composition.
For every \(x=x^{(1)}\cdots x^{(m)}\) and
\(z=z^{(1)}\cdots z^{(m)}\) in \(P^m\),
\[
 \AT(x,z)=\sum_{j=1}^{m}\AT(x^{(j)},z^{(j)}).
\]
Consequently, \(P^*\) is \(t\)-correcting if and only if \(P\) is
\(t\)-correcting, and \(m_t(P)\in\{1,\infty\}\).
\end{proposition}

\begin{proof}
Match the \(k\)-th occurrence of each symbol in \(x\) to its \(k\)-th
occurrence in \(z\), and let \(\pi(i)\) be the position matched to
position \(i\). Every swap of two unequal symbols preserves the order of equal symbols,
so a path from \(x\) to \(z\) realizes this matching. Its length is at
least the inversion count of \(\pi\), since an adjacent swap changes that
count by one. Sorting \(\pi\) by adjacent inversions attains the count;
each such swap exchanges different symbols. This gives the distance
representation in~\cite[Section~3.3]{cicirello2020kendall}.
Equal compositions of corresponding blocks also give equal cumulative
symbol counts before each block. The matching therefore stays within
corresponding block intervals, and its inversion count separates over the
blocks:
\[
 \AT(x^{(1)}\cdots x^{(m)},z^{(1)}\cdots z^{(m)})
 =\sum_{j=1}^{m}\AT(x^{(j)},z^{(j)}).
\]
The correction criterion follows: distinct words of \(P^m\) differ in some
corresponding block, so Lemma~\ref{lem:distance} and the identity make every
power \(t\)-correcting once \(P\) is, while a \(P\) that is not
\(t\)-correcting has a collision already in \(P^1=P\). Hence
\(m_t(P)\in\{1,\infty\}\).
\end{proof}

Thus a finite first collision power greater than one requires blocks with
different compositions.

\begin{remark}[Deletion-channel reduction and systematic constructions]
\label{rem:deletion-reduction}
A direct map from swap paths to common deletions relates deletion-correcting constructions to the present channel. 
This application enlarges the alphabet; the extraction below keeps it fixed.

\emph{(i) General reduction:} Along a path of \(s\) adjacent swaps, distinguish individual symbol occurrences and select one occurrence from each swapped pair. Deleting the set of selected occurrences from both endpoints leaves the same subsequence: every pair of retained occurrences keeps its relative order. At most \(s\) occurrences are deleted from each endpoint. For example, a single swap \(ab \mapsto ba\) leaves either \(a\) or \(b\) as a common subsequence. Here a symbol-invariant deletion preserves the values of the surviving symbols. Applying this reduction to a direct path of at most \(2t\) swaps between two codewords shows that a code correcting at most \(2t\) symbol-invariant deletions has adjacent-transposition distance at least \(2t+1\).

\emph{(ii) Application to explicit families:} For fixed \(t\), and message
length sufficiently large as a function of \(t\), the systematic
constructions of Wu et al.~\cite[Theorems~6 and~9(i)]{wu2026systematic},
applied with deletion budget \(2t\), insert \(8t\) markers into permutation
messages, in which each symbol of the message alphabet occurs exactly once, or \(10t\)
into messages of any prescribed composition.
In these constructions, removing the inserted markers from a codeword recovers its
message in order; this is the meaning of \emph{systematic} for these insertion-based constructions.
The deletion guarantee applies to the complete encoded words, markers included,
under the deletion model in part~(i).
The same new, distinct symbols are used once each in every codeword.
Each resulting block set \(P\) consists of the codewords of all messages in one
class, with one message length and one message multiset. Its block length is the
message length plus \(8t\) or \(10t\), its alphabet consists of the message symbols and
the \(8t\) or \(10t\) markers, and the fixed message and marker multisets give all
its blocks one composition.
The cited general construction treats message alphabets with at least two
symbols; a one-symbol message class is a singleton code.
These marker counts refer to the enlarged alphabet.
The distance bound makes each resulting block set \(P\) \(t\)-correcting;
Proposition~\ref{prop:constant-composition} then lifts correction to every
power, yielding \(m_t(P)=\infty\).
\end{remark}

\subsection{Composition residues and code size}

Proposition~\ref{prop:constant-composition} treats blocks of one fixed
composition. Congruent composition vectors also suffice when the modulus is
large enough relative to the error budget. The proof of
Theorem~\ref{thm:modular-composition} uses distance additivity only after a
hypothetical pair within distance \(2t\) has been forced to have equal
corresponding-block compositions. For such a pair, the prefix-composition difference
at each internal block boundary is componentwise a multiple of the modulus \(\mu\), so a
nonzero one would have half \(\ell_1\)-norm at least \(\mu\), and the following bound would
give distance at least \(\mu^2>2t\); hence these differences vanish, and corresponding
blocks have equal compositions. The following bound controls the
composition difference that a short sequence of swaps can create at a cut.

\begin{lemma}[A bound on prefix composition]
\label{lem:prefix-flow-amplitude}
Let \(x,z\in\Sigma_q^n\) have the same composition, and put
\[
 \Delta_j=\operatorname{comp}(x_1\cdots x_j)
          -\operatorname{comp}(z_1\cdots z_j),
 \qquad H_j=\tfrac12\lVert\Delta_j\rVert_1
 \quad(0\leq j\leq n).
\]
Then \(\AT(x,z)\geq H_a^2\) for every \(0\leq a\leq n\).
The bound is attained over two symbols, as shown in the tightness
construction for Theorem~\ref{thm:modular-composition}.
\end{lemma}

\begin{proof}
Lemma~\ref{lem:prefix-flow} gives
\(\AT(x,z)\geq\sum_{j=1}^{n-1}H_j\).
The coordinates of \(\Delta_j\) sum to zero, so \(H_j\) is a
nonnegative integer. Moreover, \(H_0=H_n=0\), and
\(\Delta_{j+1}-\Delta_j=e_{x_{j+1}}-e_{z_{j+1}}\) gives
\(\lvert H_{j+1}-H_j\rvert\leq1\). If \(H_a=h>0\), then
\(h\leq\min\{a,n-a\}\) and
\(H_j\geq\max\{h-|j-a|,0\}\). For \(1\leq r<h\), the indices
\(a\pm r\) are internal cuts and each contributes at least \(h-r\).
Consequently,
\[
 \sum_{j=1}^{n-1}H_j\geq h+2\sum_{r=1}^{h-1}r=h^2.
\]
The case \(h=0\) is immediate.
\end{proof}

\begin{theorem}[Congruent block compositions]
\label{thm:modular-composition}
Let \(t,b,\mu\geq1\) be integers with \(\mu^2>2t\). Suppose that the
nonempty set \(P\subseteq\Sigma_q^b\) satisfies
\[
 \operatorname{comp}(p)\equiv\operatorname{comp}(p')\pmod{\mu}
 \qquad\text{componentwise for all }p,p'\in P.
\]
Then \(P^*\) is \(t\)-correcting if and only if \(P\) is
\(t\)-correcting. In particular,
\(m_t(P)\in\{1,\infty\}\).
The condition \(\mu^2>2t\) is tight as a uniform guarantee: for every
\(q\geq2\) and integers \(t,\mu\geq1\) with \(\mu^2\leq2t\), there is
some block length admitting a nonempty set \(P\) whose composition
vectors are congruent modulo \(\mu\), and which is \(t\)-correcting at
the first power but not at every power.
\end{theorem}

\begin{proof}
The first power gives one direction. Conversely, suppose that \(P\) is
\(t\)-correcting and that distinct words
\(x=x^{(1)}\cdots x^{(m)}\) and \(z=z^{(1)}\cdots z^{(m)}\) in
\(P^m\) satisfy \(\AT(x,z)\leq2t\). They have equal total
composition. At every internal block boundary,
\[
 \Delta_{ib}=\sum_{j=1}^{i}
 \bigl(\operatorname{comp}(x^{(j)})-
       \operatorname{comp}(z^{(j)})\bigr)
 \in \mu\mathbb Z^{\Sigma_q}.
\]
A nonzero vector in this lattice whose coordinates sum to zero has
positive-coordinate sum at least \(\mu\). Thus \(\Delta_{ib}\ne0\)
would give \(H_{ib}\geq \mu\) and hence \(\AT(x,z)\geq \mu^2>2t\)
by Lemma~\ref{lem:prefix-flow-amplitude}. The prefix compositions at
all block boundaries therefore agree. Taking successive differences
shows that each pair of corresponding blocks has equal composition.

Match occurrences of each symbol in their left-to-right order, as in
Proposition~\ref{prop:constant-composition}. Equality of the compositions
within each pair of corresponding blocks keeps this matching inside the corresponding block
interval. Its inversions are precisely the within-block inversions, giving
\[
 \AT(x,z)=\sum_{j=1}^{m}\AT(x^{(j)},z^{(j)}).
\]
At least one pair differs, so one summand is at least \(2t+1\) by
Lemma~\ref{lem:distance}, a contradiction.

For tightness, choose distinct symbols \(\alpha,\beta\), set \(b=\mu\),
and take \(P=\{\alpha^\mu,\beta^\mu\}\). Its compositions are congruent
modulo \(\mu\), and its two blocks have infinite mutual distance. There is a
path of length \(\mu^2\) from \(\alpha^\mu\beta^\mu\) to
\(\beta^\mu\alpha^\mu\): for \(r=1,\ldots,\mu\), perform the swaps at
indices \(\mu+r-1,\mu+r-2,\ldots,r\), moving the next \(\beta\) to
position \(r\). The prefix bound with \(H_\mu=\mu\) shows that this path
is shortest. If \(\mu^2\leq2t\), its vertex after
\(\lfloor \mu^2/2\rfloor\) swaps is within
\(\lceil \mu^2/2\rceil\leq t\) swaps of both endpoints, so \(P^2\)
fails to correct \(t\) swaps.
\end{proof}

Contrapositively, a block set with \(1<m_t(P)<\infty\) has block compositions that
are not all congruent modulo any \(\mu\) with \(\mu^2>2t\).

Partitioning a given code by composition residues now yields the required
block set of the same block length. This guarantees correction
and cardinality; algorithms for selection, encoding and decoding depend on the
representation of the supplied code. The additional-delay
bound of Theorem~\ref{thm:block-decoding-delay} separately applies to the
extracted block set for known-length transmissions, and for \(q=2\)
Proposition~\ref{prop:binary-streaming-decoder} decodes its concatenations within that
delay, at a cost linear in the width of a supplied layered recognizer, which is at most
\(|P|\) for a trie of the listed set.

\begin{corollary}[The cost of correction at every power]
\label{cor:all-power-code-size}
For \(q\geq2\) and \(b,t\geq1\), let \(A_t(q,b)\) be the largest
size of a \(t\)-correcting code in \(\Sigma_q^b\), and let
\(F_t(q,b)\) be the largest size of a nonempty
\(P\subseteq\Sigma_q^b\) for which \(P^*\) is \(t\)-correcting.
With \(\mu=\lfloor\sqrt{2t}\rfloor+1\), the fixed total length \(b\)
makes the last symbol-count residue determined by the other \(q-1\),
and
\[
 \left\lceil\frac{A_t(q,b)}{\mu^{q-1}}\right\rceil
 \leq F_t(q,b)\leq A_t(q,b).
\]
More generally, every nonempty \(t\)-correcting code
\(\cD\subseteq\Sigma_q^b\) contains such a block set of size at least
\(\lceil|\cD|/\mu^{q-1}\rceil\).
This is a lower bound on the retained fraction. The tightness statement
in Theorem~\ref{thm:modular-composition} concerns the modulus threshold
as a uniform guarantee.
\end{corollary}

\begin{proof}
Partition \(\cD\) by the counts of the first \(q-1\) symbols modulo
\(\mu\). There are at most \(\mu^{q-1}\) classes, and the fixed block length
determines the remaining count modulo \(\mu\). A largest class \(P\) has
size at least \(\lceil|\cD|/\mu^{q-1}\rceil\), inherits correction from
\(\cD\), and corrects at every power by
Theorem~\ref{thm:modular-composition}. Apply this to a largest ordinary
code for the lower bound; the first power gives the upper bound.
\end{proof}

For a nonempty code \(\mathcal E\subseteq\Sigma_q^b\), write
\(r_q(\mathcal E)=b-\log_q|\mathcal E|\). The subset \(P\subseteq\cD\)
supplied by Corollary~\ref{cor:all-power-code-size} satisfies
\[
 0\leq r_q(P)-r_q(\cD)\leq(q-1)\log_q \mu,
 \qquad \mu=\lfloor\sqrt{2t}\rfloor+1,
\]
and every concatenation power of \(P\) corrects \(t\) errors. The input
code \(\cD\) is assumed to correct only at length \(b\); the extraction
adds correction under arbitrary concatenation. Taking \(\cD\) of largest
possible size shows that the smallest redundancy per block among block sets
correcting at every concatenation length exceeds the ordinary length-\(b\) optimum by
at most \((q-1)\log_q \mu\), a constant for fixed \(q,t\).
The binary case is an instance of this comparison. Kova\v{c}evi\'c's upper and lower
bounds concern codes in the whole space of binary words of length \(b\), with
codewords of any weight, and count the total displacement of the ones, which equals
the adjacent-transposition distance. With the radius \(t\) and the convention after
his equation~(54) as stated at the end of Section~\ref{sec:correcting-block-sets},
they give
\(A_t(2,b)=\Theta_t(2^b/b^t)\)
\cite[Theorem~11, Corollary~12, and equation~(54)]{kovacevic2019runlength}.
Combining this estimate with Corollary~\ref{cor:all-power-code-size} gives
\(F_t(2,b)=\Theta_t(2^b/b^t)\).
Thus the optimal per-block redundancy is \(t\log_2 b+O_t(1)\)
for fixed \(t\). The derivative family of
Section~\ref{sec:correcting-block-sets} attains this order for every
\(b\geq2t+1\); the endpoint family of
Section~\ref{sec:blockwise-decoding} does so at its stated BCH lengths
\(b=2^s+1\), with \(2t<2^s-1\) and \(2^s-1-st\geq2\).
Section~\ref{sec:correcting-block-sets} thus already gives both halves of
\(F_t(2,b)=\Theta_t(2^b/b^t)\), the upper one through the upper bound used there and
the lower one through the derivative family; the cited lower bound, with
Corollary~\ref{cor:all-power-code-size}, supplies an independent existence proof for the lower half.
For their own block sets, the derivative family of
Section~\ref{sec:correcting-block-sets} and the endpoint family of
Section~\ref{sec:blockwise-decoding} provide explicit encoding and
decoding algorithms. For an extracted block set, the correction and size guarantee
is uniform over all input codes; its algorithmic use depends on how that code is represented.

In Example~\ref{ex:composition-extraction}, the extraction retains two of
three words, and the increase in redundancy per block is
\(r_2(P)-r_2(\cD)=\log_2(3/2)<1\) bit.

\section{Conclusion and open problems}
\label{sec:open}

A fixed block set corrects \(t\) successive adjacent transpositions at
every concatenation length precisely when it does so at \(2t+1\) blocks.
The test addresses correction across block boundaries, which independent
blockwise encoding requires in addition to correction of isolated blocks. It is tight
over a \(q\)-symbol alphabet exactly when \(\binom q2\geq2t\).
The guarded encoding of a labelled digraph derived from the \v{C}ern\'y
automaton supplies block sets whose first collision occurs
at \(2t+1\) blocks.

In the binary case, the derivative family, built from BCH codes, provides a reusable block encoder
and decoder with total redundancy at most \(mt\log_2 b+O_t(m)\). This
total is \(m\) times the per-block redundancy \(t\log_2 b+O_t(1)\),
which is optimal among block sets correcting at every concatenation
length: every such block set is itself a \(t\)-correcting code of length
\(b\), so Kova\v{c}evi\'c's upper bound gives it per-block redundancy at least
\(t\log_2 b-O_t(1)\) (Section~\ref{sec:correcting-block-sets}). The linear growth of the total redundancy in
\(m\) is therefore a cost of using all concatenations of a fixed block set.
For fixed
block parameters, the encoder and decoder of the derivative family take time linear in the total length, and their
data memory is bounded in terms of \(b,t\), with \(O(\log n)\)
additional bits for framing counters.
For every block set \(P\) such that \(\cC_{mb}(P)\) is \(t\)-correcting, we prove an additional decoding delay
of at most \(2tb\) received symbols. The derivative family and the endpoint family
attain zero delay and worst-case delay \(2tb\), respectively, with the
same leading per-block redundancy.

Finally, the extraction of Section~\ref{sec:extraction} retains a constant fraction
of any ordinary correcting code when \(q,t\) are fixed. It bounds the
cost of correction under arbitrary concatenation for general alphabets.

Several questions remain. Below the alphabet threshold,
Section~\ref{sec:guarded-labelled-automata} bounds the largest finite first
collision power by \(\binom q2+1\leq D_t(q)\leq2t\), with equality on the
right when \(\binom q2=2t-1\), as in \(D_2(3)=4\); we leave open the value
of \(D_t(q)\) when \(\binom q2\leq2t-2\).
For fixed \(q\) and \(t\), the ratio of
the largest size of an ordinary length-\(b\) correcting code to the largest
size of a block set of length-\(b\) words correcting at every concatenation length
is at least \(1\) and, by the extraction, at most \(\mu^{q-1}\); we leave open whether
it tends to one as \(b\) grows.

For larger alphabets, a natural objective is an explicit family combining
the size guarantees of ordinary correcting codes with efficient blockwise
encoding and decoding under arbitrary concatenation.
The binary results also suggest studying the tradeoff between redundancy
per block and decoding delay: both extremes of the delay interval, \(0\) and \(2tb\), are
attained with logarithmic redundancy per block, and how its additive constant depends on the
delay is open.
We ask which integer worst-case delays between \(0\) and \(2tb\)
can be attained by families with per-block redundancy
\(t\log_2 b+O_t(1)\), and what the smallest additive constant in this per-block redundancy
is among families whose worst-case delay is at most \(d\).
Finally, when \(t\) grows with \(b\), the dependence of redundancy,
preparation, running time, and storage on both parameters becomes
essential. The explicit bounds here provide a starting point for that
regime.

\begingroup
\small
\bibliographystyle{plain}
\pdfbookmark[1]{References}{references}
\bibliography{references}
\endgroup
\end{document}